\documentclass[twocolumn, linenumber]{aastex631}
\usepackage[]{threeparttable}
\usepackage{tabularx}
\usepackage{amsmath}
\usepackage[caption=false]{subfig}
\usepackage[none]{hyphenat}

\newcommand{\hi}{H\thinspace{\sc i}}
\newcommand{\hii}{H\thinspace{\sc ii}}
\newcommand{\ha}{H$\alpha$}
\newcommand{\hb}{H$\beta$}
\newcommand{\hr}{H$\gamma$}
\newcommand{\hg}{H$\gamma$}
\newcommand{\hd}{H$\delta$}
\newcommand{\heps}{H$\epsilon$}

\newcommand{\hei}{He\thinspace{\sc i}}

\newcommand{\oii}{O\thinspace{\sc ii}}
\newcommand{\oiii}{O\thinspace{\sc iii}}

\newcommand{\nii}{N\thinspace{\sc ii}}
\newcommand{\sii}{S\thinspace{\sc ii}}

\newcommand{\neiii}{Ne\thinspace{\sc iii}}

\submitjournal{ApJ}

\shorttitle{Low Z Galaxies from DES}
\shortauthors{Lin et al.}

\begin{document}

\title{Low Metallicity Galaxies from the Dark Energy Survey}

\author[0000-0001-8792-3091]{Yu-Heng Lin}
\affiliation{School of Physics and Astronomy, University of Minnesota,\\ 116 Church St SE, Minneapolis, Minnesota 55455, USA}
\affiliation{Minnesota Institute for Astrophysics, University of Minnesota,\\
116 Church St SE, Minneapolis, MN 55455, USA}

\author[0000-0002-9136-8876]{Claudia Scarlata}
\affiliation{School of Physics and Astronomy, University of Minnesota,\\ 116 Church St SE, Minneapolis, Minnesota 55455, USA}
\affiliation{Minnesota Institute for Astrophysics, University of Minnesota,\\
116 Church St SE, Minneapolis, MN 55455, USA}

\author[0000-0001-7166-6035]{Vihang Mehta}
\affiliation{IPAC, Mail Code 314-6, California Institute of Technology, 1200 E. California Blvd., Pasadena CA, 91125, USA}

\author[0000-0003-0605-8732]{Evan Skillman}
\affiliation{School of Physics and Astronomy, University of Minnesota,\\ 116 Church St SE, Minneapolis, Minnesota 55455, USA}
\affiliation{Minnesota Institute for Astrophysics, University of Minnesota,\\
116 Church St SE, Minneapolis, MN 55455, USA}

\author[0000-0001-8587-218X]{Matthew Hayes}
\affiliation{Stockholm University, Department of Astronomy and Oskar Klein Centre for Cosmoparticle Physics,\\ AlbaNova University Centre, SE-10691, Stockholm, Sweden\\}

\author[0000-0001-5538-2614]{Kristen B. W. McQuinn}
\affiliation{Rutgers University, Department of Physics and Astronomy,\\
136 Frelinghuysen Road, Piscataway, NJ 08854, USA}

\author[0000-0002-1067-8558]{Lucy Fortson}
\affiliation{School of Physics and Astronomy, University of Minnesota,\\ 116 Church St SE, Minneapolis, Minnesota 55455, USA}
\affiliation{Minnesota Institute for Astrophysics, University of Minnesota,\\
116 Church St SE, Minneapolis, MN 55455, USA}

\author[0000-0003-4922-0613]{Katherine Chworowsky}
\affiliation{Department of Astronomy, University of Texas at Austin,\\
2515 Speedway Blvd, Stop C1400, Austin, Texas 78712, USA}

\author[0000-0003-1249-6392]{Leonardo Clarke}
\affiliation{Department of Physics and Astronomy, University of California, 475 Portola Plaza, Los Angeles, CA 90095, USA}

\correspondingauthor{Yu-Heng Lin}
\email{lin00025@umn.edu}



\begin{abstract}

We present a new selection of 358 blue compact dwarf galaxies (BCDs) from 5,000 square degrees in the Dark Energy Survey (DES), and the spectroscopic follow-up of a subsample of 68 objects. 
For the subsample of 34 objects with deep spectra, we measure the metallicity via the direct T$_e$ method using the auroral [\oiii]$\lambda$ 4363 emission line. 
These BCDs have an average oxygen abundance of  12+log(O/H)= 7.8, with stellar masses between 10$^7$ to 10$^8$ M$_\odot$ and specific star formation rates between $\sim$ 10$^{-9}$ to 10$^{-7}$ yr$^{-1}$.
We compare the position of our BCDs with the Mass-metallicity (M-Z) and Luminosity-metallicity (L-Z) relation derived from the Local Volume Legacy sample.
We find the scatter about the M-Z relation is smaller than the scatter about the L-Z relation. We identify a correlation between the offsets from the M-Z and L-Z relation that we suggest is due to the contribution of metal-poor inflows. 
Finally, we explore the validity of the mass-metallicity-SFR fundamental plane in the mass range probed by our galaxies. 
We find that BCDs with stellar masses smaller than $10^{8}$M$_{\odot}$  do not follow the extrapolation of the fundamental plane. This result suggests that mechanisms other than the balance between inflows and outflows may be at play in regulating the position of low mass galaxies in the M-Z-SFR space.

\end{abstract}


\section{Introduction} \label{sec:intro}
In the $\Lambda$ cold dark matter ($\Lambda$CDM)  scenario for structure formation, structures form hierarchically through mergers of smaller dark matter halos into larger ones.  Gas accretes into the halos, fueling star formation and forming the primeval, low-mass galaxies \citep{Cole_2000, Springel_2005}.  These early, low mass galaxies serve as the building blocks of galaxies. They dominate the number counts at all epochs \citep{McLeod_2021}.  Nearby low mass galaxies offer a unique perspective to expand our understanding of galaxy formation.  Observationally, the evolution of galaxies can be empirically described with the scaling relations between some basic physical properties such as the stellar mass, the star-formation rate (SFR) and the gas metal content.  For example, \citet{Skillman_1989, Pilyugin_2001, Lee_2006, Guseva_2009, Berg_2012} find that galaxies from low to high metallicity obey a luminosity-metallicity relation (L-Z relation). Later studies focused on the tighter relation between the stellar mass and the oxygen abundance known as the mass-metallicity relation \citep[M-Z relation;][]{Tremonti_2004, Berg_2012, Izotov_2015}. 

The M-Z relation shows that oxygen abundance increases with stellar mass from the low to intermediate mass systems ($\leq 10^{9.5}$ M$_\odot$), and then appears to flatten for more massive spiral galaxies ($\sim 10^{10.5}$ M$_\odot$) \citep{Tremonti_2004, Berg_2012, Izotov_2015}. Some theoretical models of galaxy formation naturally explain the M-Z relation as the result of a balance between the production of metals in the star-formation episodes, and the dilution caused by inflows of pristine gas, and the removal of metals via outflows \citep{Dayal_2013, Ma_2016, Spitoni_2017}.  In these models, outflows in low mass galaxies are very efficient at removing metal-enriched gas because of the dwarfs' shallow gravitational potential wells.  As the inflow fuels star formation, and star formation drives both the stellar mass growth and outflows, a close relation between the stellar mass, the metallicity, and the star formation rate is expected. \citet{Mannucci_2010} have shown that for galaxies  more massive than $\gtrsim 10^{9}$ M$_\odot$ these three properties are indeed correlated and this relation defines a surface in the 3D space. This relation, which is found to hold out to $z\sim 4$, is referred to in the literature as the fundamental metallicity relation \citep[FMR, e.g.,][]{Curti_2020}. At a given mass, the higher the SFR, the lower the metallicity.  For galaxies with masses below $10^9$M$_{\odot}$, the FMR is highly uncertain, due to the limited number of   well studied low mass galaxies \citep{Christensen_2012,  Belli_2013, Telford_2016}.  


For low mass galaxies, the L-Z relation and M-Z relation are intriguing at the low metallicity end.  For a given oxygen abundance, many objects show brighter B band  (or $g$ band ) luminosities.  In contrast, when compared with the more fundamental M-Z relation, the offsets are substantially reduced \citep{McQuinn_2020}.  The main reason for the offset from the L-Z relation is the higher SFR of the outliers, where young massive stars contribute a significant fraction of the luminosity \citep{McQuinn_2020}.  This effect is less pronounced in massive systems, or in the M-Z relation when the stellar masses are determined from data that sample the fainter low mass stars.
Inflow of pristine gas or a minor merger can also affect the metallicity measurements, moving galaxies both in the L-Z and in the M-Z plane \citep{Ekta_2010, Pascale_2021}.  

Despite their importance, identifying low-metallicity dwarf galaxies is challenging, primarily because of their intrinsically-low luminosities and surface brightnesses \citep[see discussion in][]{Sanchez-Almeida_2017}.  Consequently, these systems are more likely to be detected during a phase of active star formation.  During this phase, young massive stars emit a strong continuum and ionize the surrounding interstellar medium (ISM), resulting in overall blue colors, and strong emission line spectra. 
Studies have successfully identified metal-poor systems using criteria based either on spectroscopic \citep{Izotov_2012, Guseva_2017,Indahl_2021,Hirschauer_2022} or photometric  diagnostics \citep{Brown_2007, James_2015,Yang_2017, Hsyu_2018, Senchyna_2019, Kojima_2020}. In this work, we push the detection of blue compact dwarf galaxies (BCDs) to two magnitudes fainter than previous studies,  using public images from the Dark Energy Survey (DES). For a sub-sample of 34 galaxies, we present spectroscopy to confirm the redshift and accurately measure nebular oxygen abundances using the direct $T_e$ method.

The structure of the paper is as follows. In Section~\ref{sec:select}, we describe the selection method of the BCDs.  We report the spectroscopic follow-up observations and data reduction in Section~\ref{sec:data}.  In Section~\ref{sec:analysis}, we describe our measurements. We discuss the results in section~\ref{sec:discussion} and conclude in section~\ref{sec:conclusion}. Throughout the paper, we assume $\Lambda$CDM cosmology with  $H_0=$67.66 km s$^{-1}$ Mpc$^{-1}$ \citep{Planck_2018}.

\section{Sample Selection} \label{sec:select}

The Dark Energy Survey \citep[DES,][]{DES_2018} is a broadband imaging survey covering 5,000 deg$^2$ of the southern sky. 
We use \textit{g, r, i, z} catalogs and images released as part of the DES DR1 data release \citep{DES_2018}, which includes data products derived from wide-area survey observations over 345 distinct nights between August 2013 to February 2016.

We selected objects with signal--to--noise ratio greater than five in the $g,r,i$ bands and three in the $z$ band, respectively, and with \verb|EXTENDED_COADD|=3.   The \verb|EXTENDED_COADD| flag, explained in detail in \citet{DES_2018}, is used to separate galaxies from point-like stars and QSOs. This signal--to--noise ratio corresponds to a magnitude brighter than 23.7, 23.4, and 22.9, in the $g$, $r$, and $i$ filters, respectively. For comparison, the imaging data of the Sloan Digital Sky Survey have a depth of 22.13, 22.70, and 22.20 in the corresponding filters \citep{SDSS_imagedata}.   

To select BCD galaxies, we used similar color selection criteria as \citet{Hsyu_2018}. 
The criteria were optimized based on the photometric colors of known low metallicity systems such as I~Zwicky 18 \citep{Zwicky_1966, Skillman_1993} and Leo~P \citep{Giovanelli_2013, Skillman_2013}. The \citet{Hsyu_2018} selection identifies blue compact sources with colors consistent with \hii\ regions. This is because  by enforcing $g$ band magnitudes brighter than $z$ band, redder host galaxies are naturally  excluded.

The color criteria are optimized to search for galaxies with colors consistent with low redshift ($z \sim  0.01 - 0.02$) low-metallicity galaxies such as I~Zwicky 18 and Leo~P. Thus, we expect to select sources at low redshift ($z \sim 0.02$), although by pushing to fainter magnitudes than the SDSS data we are able to search 
{\bf for fainter galaxies, which could be intrinsically faint or at higher redshift.} 
We note that most of our spectra were observed with the Blue (b2k) disperser, which has the shortest observed wavelength of 3795\AA. To cover the [\oii]$\lambda$3727 emission line, we only obtain the spectroscopic follow-up for objects with redshift $z\geq0.02$ in this work.

The broadband color selection from the following query resulted in 895 galaxies from the full 5,000 square degrees area. The selection criteria we used are: 
\begin{equation}
\begin{aligned}
    -0.2 \le &~g-r \le 0.2 \\
    -0.7 \le &~r -i \le -0.1 \\
    -0.4 - 2\times z_{err} \le &~i - z \le 0.1,
\end{aligned}
\end{equation}
where $z_{err}$ is the uncertainty of the $z$ band magnitude.  The complete SQL query was run on the NOAO datalab interface \footnote{\href{https://datalab.noirlab.edu/}{https://datalab.noirlab.edu/}}, and can be found in Appendix \ref{apd:SQL}. The photometric errors were taken into consideration to maximize the purity of the selected sample. {\bf We further applied a Galactic extinction cut of E(B-V)$<0.066$.}

The original criteria from \citet{Hsyu_2018}, developed for SDSS, included an additional color condition based on the $u$ band photometry, which is not available over the DES area. Therefore, we used data from the all-sky GALEX survey and transformed the $u$ band condition from \citet{Hsyu_2018} into a condition on the GALEX NUV filter \citep[centered at 2,700\AA;][]{GALEX_2005}.
We cross-matched the \citet{Hsyu_2018} objects with the GALEX survey  sources using a search aperture of 2.5\arcsec radius.  We found that 60 out of the 88 \citet{Hsyu_2018} objects also have GALEX data.  We performed an orthogonal distance regression (taking into consideration the  errors in both bands) to find a linear relation between the magnitudes in the $u$ and $NUV$ filters. The best fit linear relation is: 

\begin{equation}
    u = 0.96~\times NUV + ( 0.46 \pm 1.5 ).
\end{equation}

Using this relation, we transform the \citet{Hsyu_2018} $u$-band-based color criterion as a function of the $NUV$ magnitude as:

\begin{equation}
    -1.76 \le 0.96\times NUV - g \le 1.64
\end{equation}

\begin{figure}
\centering
\includegraphics[width=0.99\linewidth]{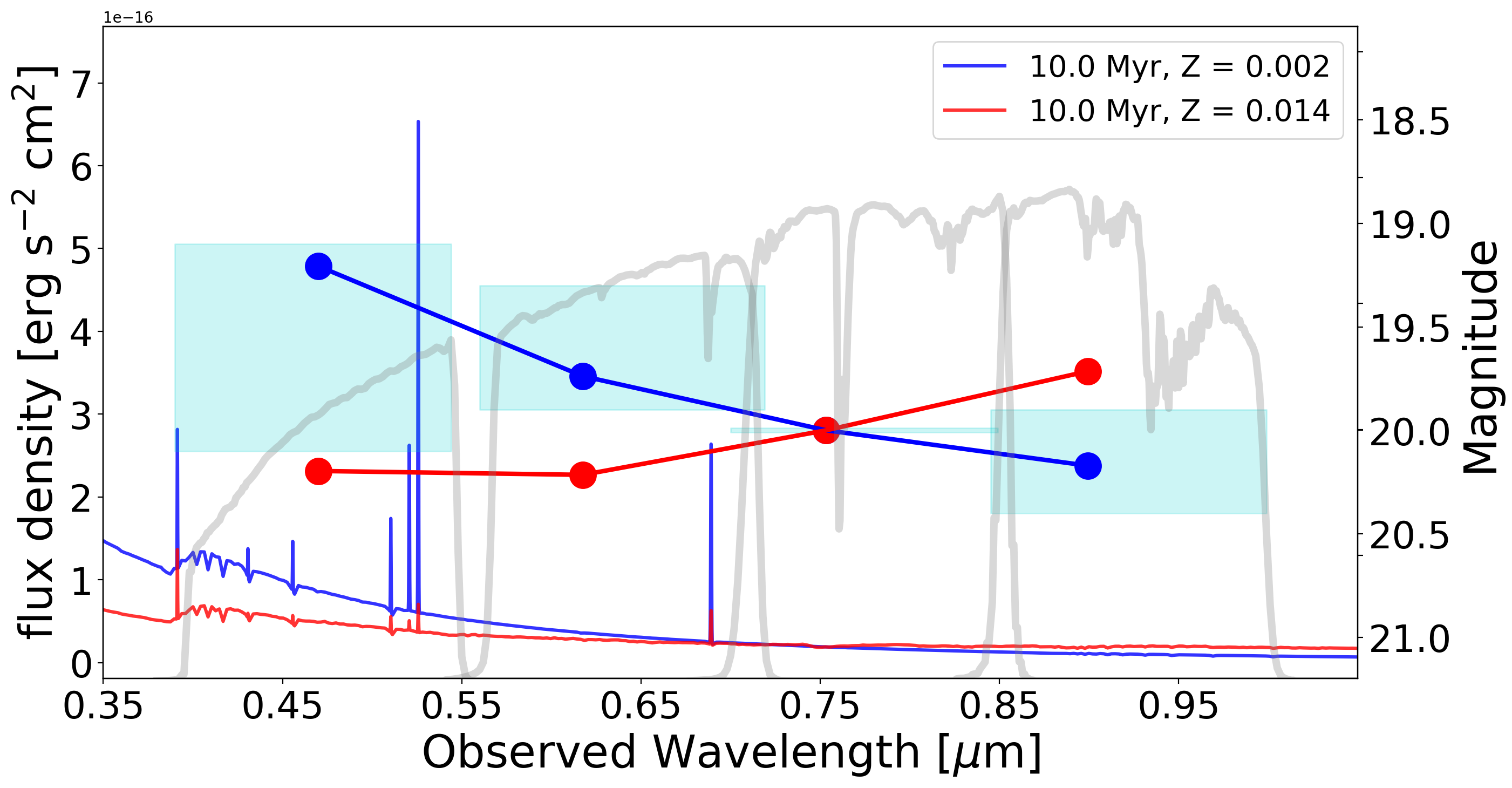}
\caption{The SED and the spectra of two stellar population models at 10 Myr. The lower metallicity model is shown in blue and the higher metallicity model is shown in red. Both spectra are normalized at $i$ magnitude=20.  The throughput of the $g, r, i, z$ filters are shown in gray. The lightblue shaded areas indicate our selection criteria. }
\label{fig:SED}
\end{figure}

In Figure~\ref{fig:SED}, we demonstrate this selection method using two stellar population synthesis models generated with  \textsc{Starburst99} \citep{Leitherer_1999}. Specifically, we consider a low metallicity model (Z=0.002, 12+log(O/H) $\simeq$ 7.83) and a high (Z=0.014, 12+log(O/H) $\simeq$ 8.67) metallicity model using the ``Geneva track without rotation" \citep{Ekstrom_2012}, both 10~Myr old, redshifted to $z = 0.05$, and normalized to $i$ magnitude $= 20$.  Fluxes for the brightest recombination lines, including \ha, \hb, \hg, are calculated based on Case~B line ratios and the intrinsic \hb\ line luminosity produced by the model. The nebular emission lines are faint 10~Myr after the star formation. We assume the gas metallicity is the same as the stellar metallicity, and estimate the oxygen line ([\oii]$\lambda\lambda$ 3726, 3729 doublets and [\oiii]$\lambda\lambda$ 4959, 5007 doublets) fluxes using the \cite{Maiolino_2008} strong line calibration. The lower metallicity model (blue line) shows a bluer stellar continuum and brighter emission lines compared to the high metallicty model (red line), resulting in an overall bluer spectral energy distribution. 
As a result, our technique successfully identifies the low metallicity galaxies.    

\begin{figure}
\centering
\includegraphics[width=0.99\linewidth]{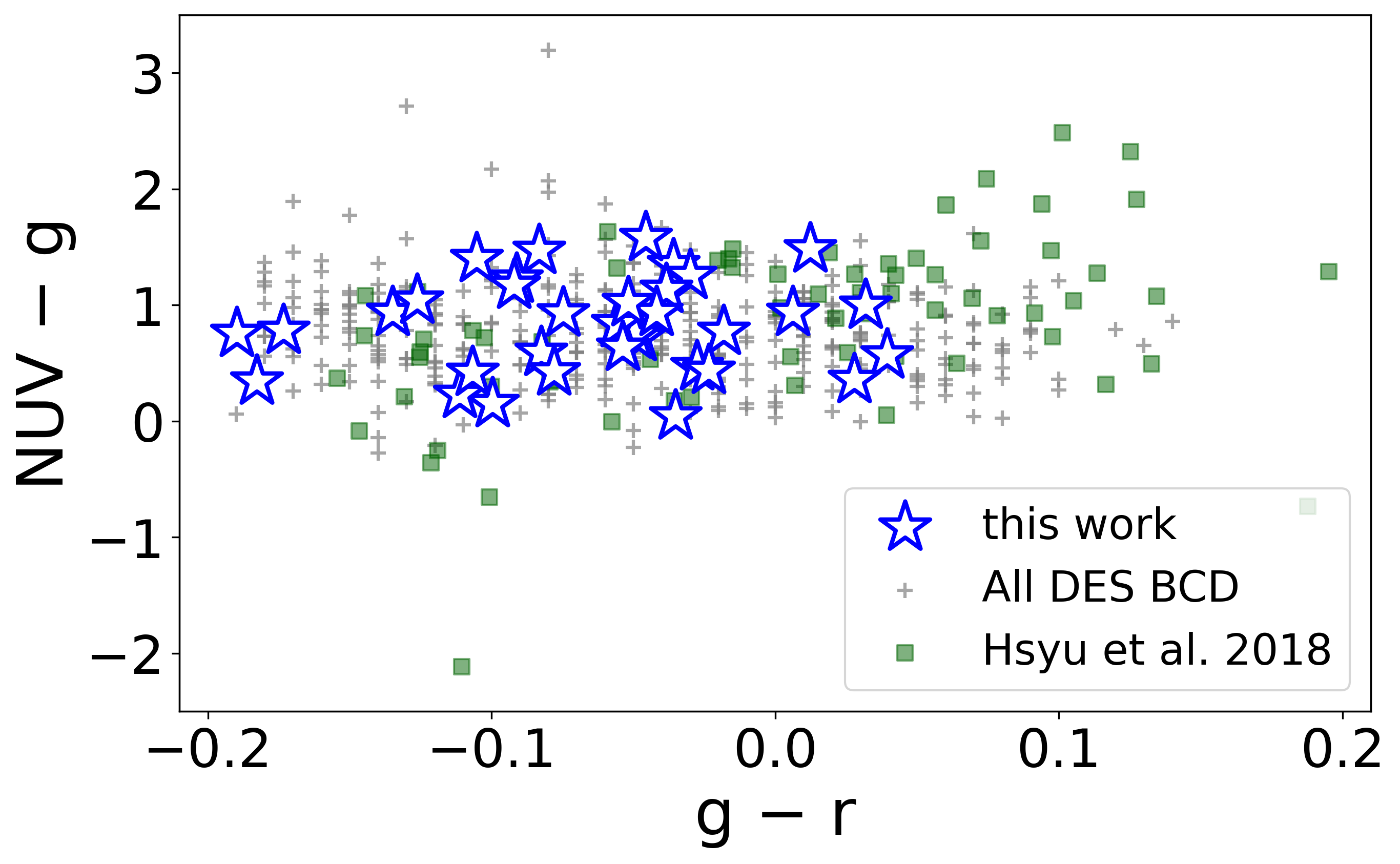}
\caption{The 358 sources chosen in this paper (gray cross) as well as \citet{Hsyu_2018} BCDs (green square). The blue stars represent sources with follow-up spectroscopic confirmation.}
\label{fig:nuv_colorcolor}
\end{figure}

Out of the 895 galaxies selected on the basis of the optical colors alone, 516 objects had a corresponding GALEX counterpart. In the following we only consider galaxies that have a GALEX detection. We created a Zooniverse Panoptes project\footnote{\href{www.zooniverse.org/lab}{www.zooniverse.org/lab}} for the authors to visually inspect all galaxies with a GALEX counterpart. In order to exclude objects where the colors were due to image artifacts, we visually examined the individual band DES images ($g$, $r$, $i$) that entered the color selection. Each object was inspected by at least 3 group members, and we kept in our final clean catalog  those 358 BCDs, for which all the experts agreed that the images were clean from artifacts. Figure~\ref{fig:nuv_colorcolor} shows the distribution in the (\textit{NUV-g})--(\textit{g-r}) color-color space of the final DES BCD galaxy sample as well as the sample of \citet{Hsyu_2018}.

\section{Observations and Data Reduction} \label{sec:data}

\subsection{Spectroscopic follow-up}
Spectroscopic observations of 68 candidates were obtained using the Cerro Tololo Ohio State Multi-Object Spectrograph (COSMOS) mounted on the 4m Blanco telescope at the CTIO facility. The observations were taken during two observing runs, a first in January 2020 and a second in  June 2021.
Targets were chosen to be visible at airmass lower than 1.3 during the observations. To optimize the observing efficiency, we prioritized observing bright objects ($g$ $\le$ 20). 
For each object, we first obtained short exposures (200 seconds) optimized for the detection of bright lines and to confirm the galaxy's redshift. We then observed the confirmed galaxies with deeper exposure times aiming at the detection of the auroral [\oiii] line required for the direct measurement of the gas-phase metallicity. For each object, we took three exposures to remove the cosmic rays. The total exposure time for each object ranged between 3$\times$600 to 3$\times$1200 seconds.

We used the Blue (b2k) disperser with a 0\farcs9 slit, covering the wavelength range between 3795\AA\ and 6615\AA, at a resolution of R=2400.  
We also observed 5 galaxies using the Wide (l2k) disperser with a 0\farcs9 slit, covering the wavelength range from 3570\AA\ to 7120 \AA, at a resolution of R=1800. The seeing in the 2020 run varied from 0\farcs7 to 0\farcs9, with sporadic high cloud coverage. The seeing in the 2021 run varied from 0\farcs8 to 1\farcs4, with medium cloud coverage. For flux calibration, we observed the standard stars   EG21 and LTT3864 in the 2020 observing run, and  LTT4816 and LTT7379 in 2021  run \citep{Stone_1983}. 

We used two different acquisition strategies in the two observing runs. In 2020, to properly place the target into the slit, we first centered on a bright star within two arcminutes from the target galaxy, and then the slit was rotated to include both the target and the acquisition star. Because the Blanco telescope does not have an atmospheric dispersion corrector (ADC), this strategy requires a correction in post processing for the wavelength-dependent flux-loss resulting from atmospheric refraction (see Section~\ref{sec:data_reduction} and Appendix~\ref{apd:slitloss}).  In 2021, we instead acquired directly on the targets (which therefore had to be slightly brighter than in the 2020 run) and oriented the slit at the parallactic angle to avoid slit-losses. During the 2021 observations, three objects selected from SDSS (labeled in Table~\ref{tab:mag}) were observed as a filling for the first half of the night. The selection criteria of these objects are the same as described in \citet{Hsyu_2018}.

\subsection{CTIO Data Reduction}\label{sec:data_reduction}

We use the \textsc{IRAF} package \citep{IRAF} to reduce and calibrate the spectroscopic data.  The reduction and calibration processes include overscan and bias subtraction, flat-fielding, wavelength calibration, background subtraction, atmospheric extinction correction, aperture extraction, and flux calibration. The wavelengths are calibrated with a HgNe lamp. The aperture sizes in the cross-dispersion direction were chosen to optimize the S/N ratio of the continuum, after verifying that the emission lines showed the same spatial distribution as the continuum.

 For the 2020 observation run, we correct the wavelength dependent slit-loss following \citet[][see details in Appendix~\ref{apd:slitloss}]{Filippenko_1982}. We assume the source has a Gaussian spatial profile with standard deviation equal to half of the seeing full width half maximum (FWHM), and calculate the flux recovery factor $\epsilon(\lambda) = I(\lambda)/I^{\prime}(\lambda)$ for both the standard star and targets, where $I^{\prime}(\lambda)$ is the observed amount of light, and $I(\lambda)$ is the amount that goes through the slit if there were no refraction. See details in Appendix~\ref{apd:slitloss}.
 
 The atmospheric refraction is most significant at the blue side of the spectra, which affects the flux measurements of [\oii]$\lambda$3727 and the derived 12+ log(O/H). The mean and maximum of the flux recovery factor $\epsilon(3727)$ are 1.36 and 1.66. Uncorrected, this flux loss would lead to a 0.13 and 0.22 dex lower values in the calculated oxygen abundances. In the 2021 observation run, we observed at the parallactic angle, so we do not apply a flux correction.  

\begin{figure*}[ht]
\centering
\subfloat[]{%
  \includegraphics[clip,width=0.9\textwidth]{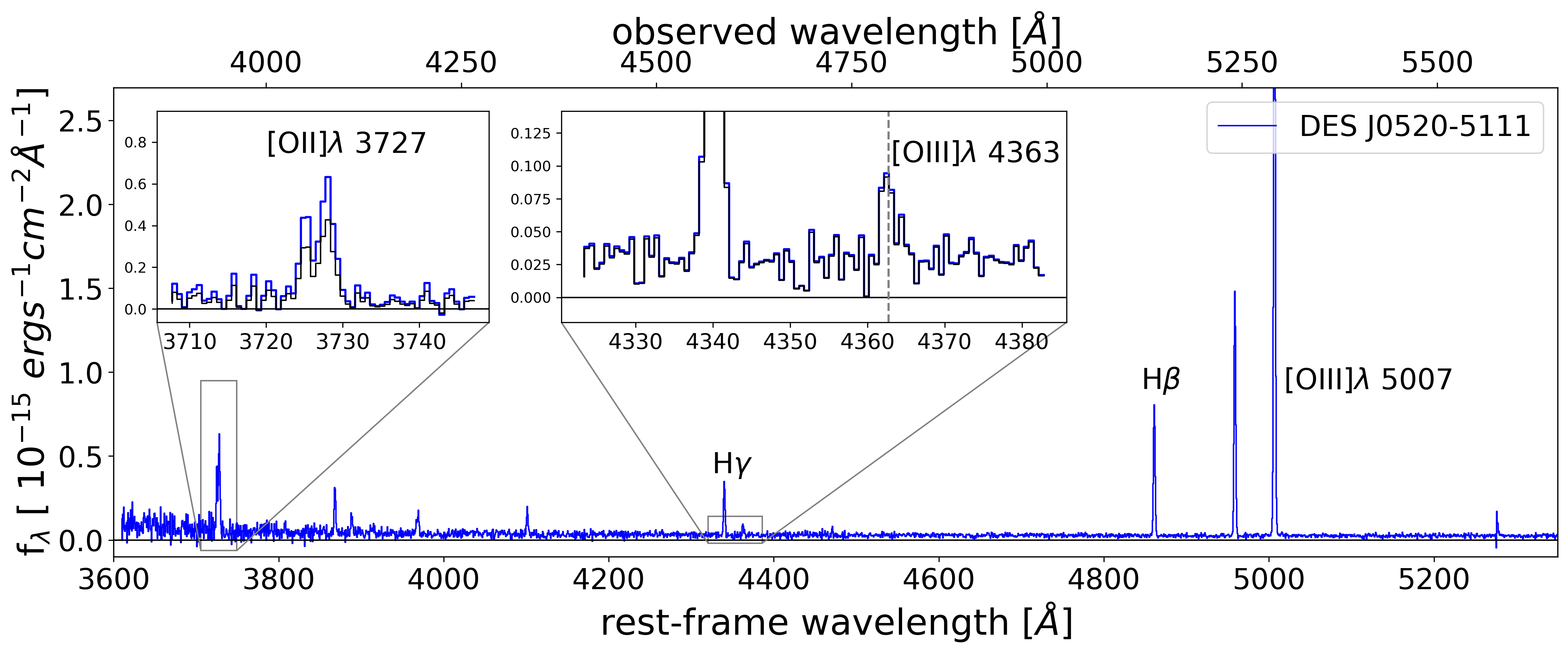}%
  }
  \\
\subfloat[]{%
  \includegraphics[clip,width=0.9\textwidth]{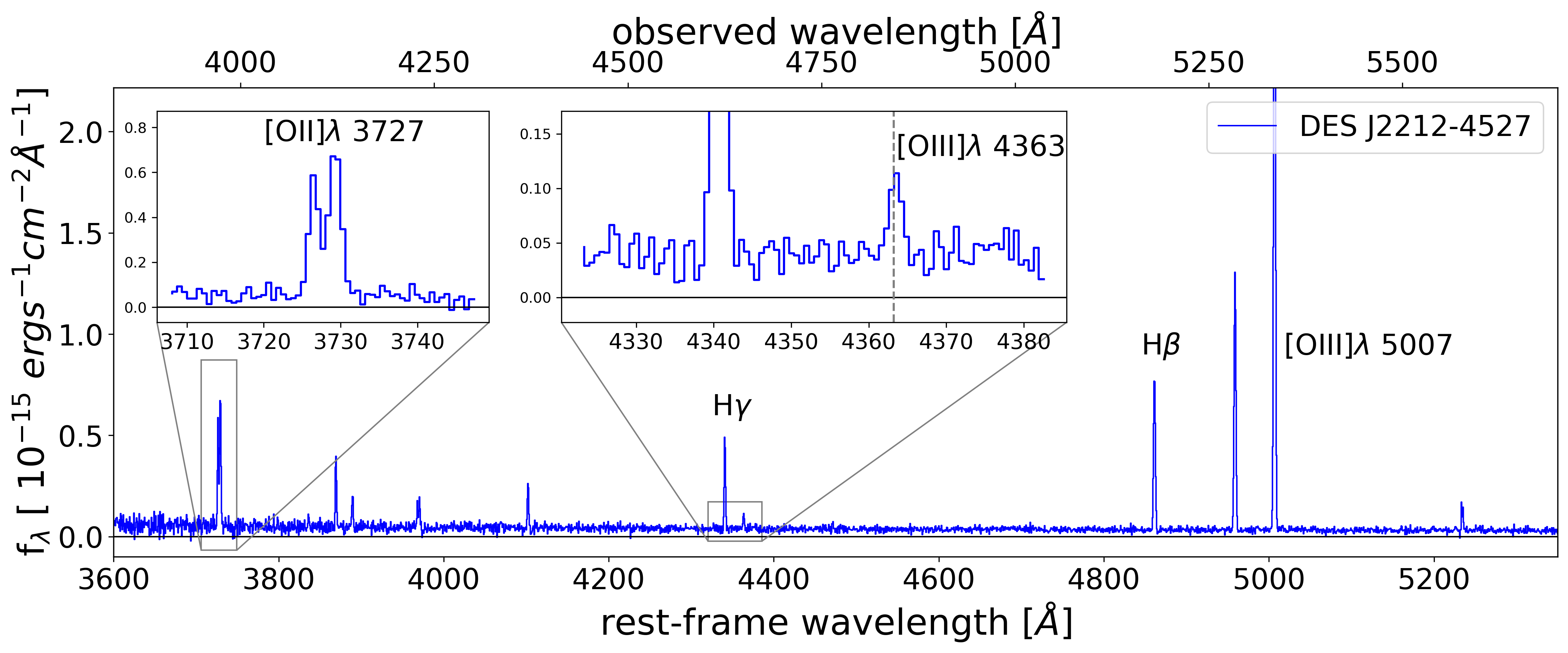}%
}
 \\
\subfloat[]{%
  \includegraphics[clip,width=0.9\textwidth]{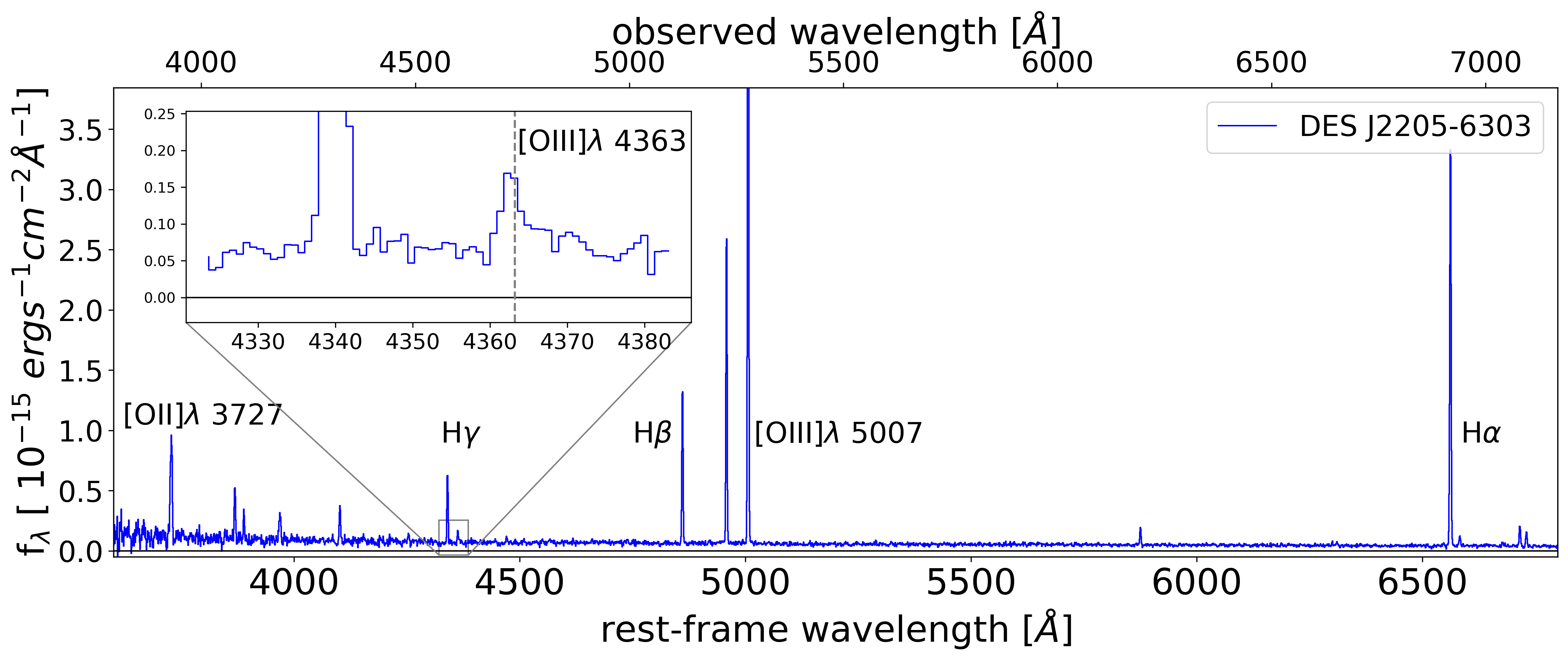}%
}

\caption{Three example spectra of BCDs. (a)The b2k spectrum observed in Jan., 2020. The intensity before (black) and after (blue) the atmospheric refraction correction are shown in the inset  zooming-in on the [\oii]$\lambda$3727 and [\oiii]$\lambda$4363 lines. (b) The b2k spectrum observed in June, 2021.  (c) The l2k spectrum observed in June, 2021.}
\label{fig:spec}
\end{figure*}

Figure~\ref{fig:spec}(a) shows an example of the b2k spectrum taken in 2020, with the insets zooming-in on the [\oii]$\lambda$ 3727 doublet and the temperature sensitive [\oiii]$\lambda$4363 emission line.  Figure~\ref{fig:spec}(b) and Figure~\ref{fig:spec}(c) shows the b2k and l2k spectra taken in 2021. 

Using the short exposures, we confirm that 59 out of the 67 BCD candidates are emission line galaxies,  yielding a $\sim$ 90\% success rate for the chosen selection criteria. For the 35 galaxies with deep spectroscopic follow-up, we report the redshifts together with the {\bf Galactic extiction-corrected} photometry in Table~\ref{tab:mag}, and the 59 confirmed objects in the online material. For all 34 galaxies (14 from the 2020, and 20 from the 2021 run) we were able to detect the auroral [\oiii]$\lambda$4363 line at a signal--to--noise ratio $>$5.

%
%
%

\begin{table*} 
\centering 
\caption{The coordinates, the measured spectroscopic redshift, the magnitudes in FUV, NUV, $g$, $r$, $i$, $z$ bands of the BCDs presented in this work. The magnitudes are corrected for Galactic extinction.  The full sample with spectroscopic confirmed redshift are included in the online material.  a: The last three objects are selected from SDSS.  } 
\begin{threeparttable} 
\begin{tabular*}{1.05\textwidth}{cccccccccc} 
 ID & RA & Dec & redshift & FUV & NUV & $g$ & $r$ & $i$ & $z$ \\ \hline\hline 
  J0006-5911 & 0:06:52.47 & -59:11:13.01 & 0.0109 & 20.16$\pm$0.16 & 20.00$\pm$0.10 &19.43$\pm$0.01 & 19.39$\pm$0.01 &19.67$\pm$0.02  & 19.64$\pm$0.03  \\ 
 J0019-5937 & 0:19:15.17 & -59:37:16.41 & 0.0357 & 19.52$\pm$0.17 & 19.31$\pm$0.09 &18.38$\pm$0.01 & 18.38$\pm$0.01 &18.50$\pm$0.01  & 18.44$\pm$0.01  \\ 
 J0212-3231 & 2:12:58.59 & -32:31:04.48 & 0.0490 & 20.50$\pm$0.24 & 20.08$\pm$0.12 &19.33$\pm$0.01 & 19.52$\pm$0.01 &19.70$\pm$0.01  & 19.67$\pm$0.02  \\ 
 J0218-1946 & 2:18:05.51 & -19:46:07.85 & 0.0304 & 21.23$\pm$0.29 & 21.33$\pm$0.23 &19.87$\pm$0.01 & 19.95$\pm$0.01 &20.17$\pm$0.02  & 20.08$\pm$0.03  \\ 
 J0218-0912 & 2:18:52.90 & -9:12:18.28 & 0.0127 & 19.94$\pm$0.06 & 19.68$\pm$0.04 &18.35$\pm$0.01 & 18.39$\pm$0.01 &18.53$\pm$0.01  & 18.46$\pm$0.01  \\ 
 J0242-5149 & 2:42:01.53 & -51:49:37.08 & 0.0526 & 20.66$\pm$0.19 & 21.52$\pm$0.24 &20.28$\pm$0.01 & 20.31$\pm$0.01 &20.49$\pm$0.03  & 20.50$\pm$0.05  \\ 
 J0249-0111 & 2:49:39.72 & -1:11:51.01 & 0.0672 & -- & 19.41$\pm$0.04 &18.52$\pm$0.01 & 18.65$\pm$0.01 &18.80$\pm$0.01  & 18.81$\pm$0.01  \\ 
 J0311-1341 & 3:11:44.15 & -13:41:36.36 & 0.0311 & 20.68$\pm$0.09 & 20.49$\pm$0.07 &19.80$\pm$0.01 & 19.85$\pm$0.01 &20.14$\pm$0.02  & 20.13$\pm$0.03  \\ 
 J0426-5410 & 4:26:03.93 & -54:10:50.03 & 0.0551 & -- & 20.18$\pm$0.17 &19.18$\pm$0.01 & 19.15$\pm$0.01 &19.27$\pm$0.01  & 19.26$\pm$0.01  \\ 
 J0450-5429 & 4:50:38.40 & -54:29:27.38 & 0.0471 & 18.99$\pm$0.08 & 19.05$\pm$0.07 &18.91$\pm$0.01 & 19.01$\pm$0.01 &19.20$\pm$0.01  & 19.17$\pm$0.01  \\ 
 J0500-3204 & 5:00:24.60 & -32:04:11.06 & 0.0401 & 20.61$\pm$0.36 & 20.96$\pm$0.35 &19.49$\pm$0.01 & 19.48$\pm$0.01 &19.65$\pm$0.02  & 19.57$\pm$0.03  \\ 
 J0520-5111 & 5:20:04.06 & -51:11:37.80 & 0.0569 & 20.90$\pm$0.31 & 20.72$\pm$0.20 &19.60$\pm$0.01 & 19.64$\pm$0.01 &19.82$\pm$0.01  & 19.76$\pm$0.02  \\ 
 J0522-2251 & 5:22:37.37 & -22:51:31.82 & 0.0310 & -- & 20.46$\pm$0.07 &20.14$\pm$0.01 & 20.32$\pm$0.02 &20.84$\pm$0.04  & 20.99$\pm$0.09  \\ 
 J0524-2041 & 5:24:24.16 & -20:41:19.05 & 0.0093 & 20.84$\pm$0.30 & 20.73$\pm$0.23 &19.37$\pm$0.01 & 19.47$\pm$0.01 &19.76$\pm$0.02  & 19.77$\pm$0.04  \\ 
 J0527-2704 & 5:27:05.68 & -27:04:06.74 & 0.0534 & 19.67$\pm$0.13 & 20.02$\pm$0.11 &18.99$\pm$0.01 & 19.12$\pm$0.01 &19.31$\pm$0.01  & 19.24$\pm$0.02  \\ 
 J0617-4954 & 6:17:52.66 & -49:54:50.88 & 0.0512 & -- & 20.97$\pm$0.12 &20.22$\pm$0.01 & 20.39$\pm$0.02 &20.69$\pm$0.03  & 20.71$\pm$0.06  \\ 
 J2014-5348 & 20:14:23.68 & -53:48:03.51 & 0.0584 & 21.55$\pm$0.35 & 20.65$\pm$0.19 &20.21$\pm$0.01 & 20.24$\pm$0.01 &20.47$\pm$0.03  & 20.50$\pm$0.04  \\ 
 J2030-4719 & 20:30:54.90 & -47:19:41.70 & 0.0478 & 20.12$\pm$0.15 & 19.62$\pm$0.10 &19.42$\pm$0.01 & 19.53$\pm$0.01 &19.76$\pm$0.01  & 19.76$\pm$0.02  \\ 
 J2036-4918 & 20:36:09.86 & -49:18:38.23 & 0.0251 & 19.87$\pm$0.13 & 20.01$\pm$0.11 &19.67$\pm$0.01 & 19.64$\pm$0.01 &19.83$\pm$0.02  & 19.81$\pm$0.03  \\ 
 J2044-5731 & 20:44:02.95 & -57:31:02.15 & 0.0249 & 19.70$\pm$0.22 & 19.96$\pm$0.16 &19.22$\pm$0.01 & 19.24$\pm$0.01 &19.45$\pm$0.01  & 19.42$\pm$0.02  \\ 
 J2051-4559 & 20:51:37.74 & -45:59:24.60 & 0.0499 & 20.17$\pm$0.22 & 20.15$\pm$0.12 &19.58$\pm$0.01 & 19.67$\pm$0.01 &19.91$\pm$0.02  & 19.98$\pm$0.05  \\ 
 J2052-4517 & 20:52:40.45 & -45:17:47.10 & 0.0309 & 19.98$\pm$0.22 & 20.21$\pm$0.14 &19.02$\pm$0.01 & 19.11$\pm$0.01 &19.31$\pm$0.01  & 19.31$\pm$0.02  \\ 
 J2055-5405 & 20:55:56.73 & -54:05:59.11 & 0.0622 & 20.29$\pm$0.15 & 20.27$\pm$0.12 &19.56$\pm$0.01 & 19.61$\pm$0.01 &19.86$\pm$0.01  & 19.87$\pm$0.03  \\ 
 J2109-5228 & 21:09:53.43 & -52:28:36.54 & 0.0498 & 19.51$\pm$0.11 & 19.46$\pm$0.08 &18.64$\pm$0.01 & 18.69$\pm$0.01 &18.86$\pm$0.01  & 18.81$\pm$0.02  \\ 
 J2117-5920 & 21:17:49.95 & -59:20:30.91 & 0.0582 & 20.63$\pm$0.20 & 20.42$\pm$0.14 &19.51$\pm$0.01 & 19.55$\pm$0.01 &19.73$\pm$0.02  & 19.73$\pm$0.03  \\ 
 J2120-5054 & 21:20:09.21 & -50:54:30.18 & 0.0434 & 19.23$\pm$0.11 & 19.08$\pm$0.06 &19.05$\pm$0.01 & 19.09$\pm$0.01 &19.25$\pm$0.01  & 19.18$\pm$0.01  \\ 
 J2203-4013 & 22:03:03.95 & -40:13:03.53 & 0.0366 & -- & 19.99$\pm$0.14 &19.58$\pm$0.01 & 19.69$\pm$0.01 &19.99$\pm$0.02  & 19.94$\pm$0.03  \\ 
 J2205-6303 & 22:05:04.95 & -63:03:28.80 & 0.0542 & 19.96$\pm$0.22 & 20.24$\pm$0.21 &19.33$\pm$0.01 & 19.40$\pm$0.01 &19.57$\pm$0.01  & 19.57$\pm$0.02  \\ 
 J2205-4218 & 22:05:40.78 & -42:18:25.25 & 0.0270 & 20.39$\pm$0.11 & 20.01$\pm$0.07 &19.58$\pm$0.01 & 19.61$\pm$0.01 &19.77$\pm$0.01  & 19.71$\pm$0.02  \\ 
 J2212-4527 & 22:12:34.77 & -45:27:36.89 & 0.0656 & -- & 21.00$\pm$0.27 &19.84$\pm$0.01 & 19.94$\pm$0.01 &20.12$\pm$0.02  & 20.16$\pm$0.03  \\ 
 J2213-4320 & 22:13:31.83 & -43:20:38.16 & 0.0679 & 20.57$\pm$0.17 & 19.99$\pm$0.09 &19.58$\pm$0.01 & 19.66$\pm$0.01 &19.82$\pm$0.01  & 19.81$\pm$0.02  \\ 
 J2237-4845 & 22:37:24.75 & -48:45:28.14 & 0.0502 & 19.32$\pm$0.08 & 19.12$\pm$0.05 &18.49$\pm$0.01 & 18.55$\pm$0.01 &18.71$\pm$0.01  & 18.71$\pm$0.01  \\ 
 J1403+0151 & 14:03:34.29 & +1:51:17.55 & 0.0243 & -- & -- &19.54$\pm$0.04 & 19.58$\pm$0.07 &20.27$\pm$0.18  & 20.36$\pm$0.79  \\ 
 J1409+0116 & 14:09:47.30 & +1:16:11.43 & 0.0254 & 19.93$\pm$0.04 & 19.95$\pm$0.04 &18.39$\pm$0.02 & 18.44$\pm$0.03 &18.58$\pm$0.05  & 18.88$\pm$0.24  \\ 
 J1451+0231 & 14:51:07.12 & +2:31:25.03 & 0.0068 & 18.52$\pm$0.02 & 18.77$\pm$0.04 &17.78$\pm$0.01 & 17.83$\pm$0.02 &18.02$\pm$0.03  & 18.13$\pm$0.12  \\ 
\hline
\end{tabular*} 
\label{tab:mag} 
\end{threeparttable}
\end{table*}

\section{Analysis}\label{sec:analysis}

In this section, we first explain the measurements of the emission line fluxes (Section \ref{sec:lineflux}), and then the steps involved in the measurement of the galaxies' physical properties, including the gas oxygen abundance  (Section \ref{sec:metallcity}),  star formation rate (Section~\ref{sec:SFR}), and  stellar mass (Section~\ref{sec:mass}).

\subsection{Emission line measurements}\label{sec:lineflux}

We estimate the redshift $z$ using the brightest forbidden emission line [\oiii ] $\lambda$5007 \AA.  
For each of the emission lines ($i$) we project the spectrum to  velocity space,
and measure the continuum level as the median of the flux density in the velocity range of $v$ $\pm$ 500 km s$^{-1} -$ 1000 km s$^{-1}$, after masking other emission lines within this range.  The emission line flux is measured by direct integration of the spectral flux density over the continuum-subtracted line profile, typically within $\mid v \mid \le$ 200 km s$^{-1}$.  
The uncertainty of the emission line flux is calculated from the standard deviation of the continuum, adding in quadrature a  2$\%$ systematic flux uncertainty resulting from the flux calibration step. Emission line measurements for the sample galaxies are reported in Table~\ref{tab:emission}.

We correct for Galactic extinction using the dust maps in \citet{SFD} and the reddening curve of \citet{Cardelli_1989} with $R_V$ = 3.1. 
\subsection{Electron temperature and internal dust extinction correction}\label{sec:Te}

In addition to Galactic extinction we also account for internal extinction, using the emissivity of the Balmer lines under the Case~B assumption. Because the intrinsic Balmer line ratios depend mildly on the electron density, n$_e$, and the electron temperature, T$_e$, we proceed with the following iterative approach.  We assume a fixed electron density n$_e$=100 cm$^{-3}$ throughout the calculation of the dust extinction correction.  We  calculate the O$^{++}$ electron temperature T$_e$(\oiii) using the line ratio of [\oiii ]$\lambda\lambda$4959,5007/[\oiii ]$\lambda$ 4363 with the method described in \citet{Izotov_2006}, and the O$^{+}$ electron temperature T$_e$(\oii) with the empirical relation \citep{Pagel_1992}: 

\begin{equation}
    t_e(\text{\oii}) = \frac{2}{t_e(\text{\oiii})^{-1} + 0.8},
\end{equation}

\noindent
where $t_e$ = T$_e\times 10^{-4}$. 
The first estimate of the temperature is calculated without the dust reddening correction. Once we obtain the color excesses E(B$-$V), we correct the emission line fluxes and repeat the calculation described above until T$_e$(\oiii) changes by less than 1000K.  We constrain our temperature range to be between 5000K $\le$ T$_e$(\oiii) $\le$ 22000K, since the \hii\ regions rarely exceed this temperature, and the overestimation of T$_e$(\oiii) leads to an underestimation of the oxygen abundance.


We correct for the dust reddening and the underlying stellar absorption using the minimum $\chi^2$ approach\footnote{If the uncertainties of the line fluxes are normally distributed, then the ratios of emission lines are described by the Lorentz distribution (also known as Cauchy distribution). However, our approach is a good approximation given the S/N of the emission lines \citep{Wesson_2016}. }: 

\begin{equation}
    \chi^2 = \sum_{\lambda} \frac{(X_{obs}(\lambda) - X_{ext}(\lambda))^2}{\sigma^2_{X_{obs}}(\lambda)}, 
\end{equation}
where $X_R(\lambda)$, $\sigma^2_{X_{obs}}$ are the observed Balmer line ratios and uncertainties over \hb. $X_{ext}(\lambda) = F_\lambda/$\hb\ is the product of extinction and intrinsic line ratio under Case~B assumption.  The extinction at a given wavelength $\lambda$ is computed as the following: 

\begin{equation}
    F_\lambda = F_\lambda^0~\frac{10^{-0.4~R_V~f(\lambda)~E(B-V)} } 
    { ( 1 +  \frac{ k(\lambda)a_{\text{\hi}}} {\text{EW}(\lambda)}) }
\end{equation}

where $F_\lambda^0$ is the flux without extinction, $f(\lambda)$ is the reddening function from \citet{Cardelli_1989};  R$_V$ $\equiv$ A$_V$/E(B$-$V) = 3.1.  $EW(\lambda)$ is the equivalent width of the line, and $a_{\text{\hi}}$ ($\leq$ 0) is the equivalent width of the underlying stellar absorption at \hb; $k(\lambda)$ is the scaling coefficient accounting for the wavelength dependence of the underlying stellar absorption \citep{Aver_2021}.  We use  \textsc{PyNeb} \citep{pyneb} to calculate the theoretical line ratios under the Case~B assumption, as a function of density and temperature. We include the \ha\ (if present), \hb, \hg, and \hd\ in our calculation, while \heps\ and H8 are not included due to the blends of [\neiii ] and \hei, respectively, and we do not have high S/N detections on H9 in many spectra.  

The dust-extinction-corrected emission line fluxes are presented in Table~\ref{tab:emission}, and the T$_e$ and derived oxygen abundances are presented in Table~\ref{tab:property}.  The uncertainties of the dust extinction and the T$_e$ are computed via a Monte Carlo simulation.  We generate 500 realizations by randomizing the Balmer line ratios within $\pm 1\sigma$.  For each realization, we compute the E(B-V) and the T$_e$. We report the 16$\%$ and 84$\%$ percentiles of the distribution.

\subsection{Metallicity}\label{sec:metallcity}

We use the total oxygen abundance relative to hydrogen as a proxy of the  \hii\ region metallicity.
We also assume that the oxygen total abundance can be written as: 

\begin{equation}
    \frac{\text{O}}{\text{H} } = \frac{\text{O}^+}{\text{H}^+} +\frac{\text{O}^{++}}{\text{H}^+}.
\end{equation}

The oxygen abundance can be determined directly when the electron density, temperature, and line emissivities are known. This method is referred to as the direct method in the literature.  When the gas metallicity is estimated empirically, using only the strongest emission lines, the measurements are said to use the strong line calibration. We summarize the direct method in Section~\ref{sec:direct} \citep[see, e.g.,][]{Izotov_2006}.  In Section~\ref{sec:SL}, we also consider the alternative empirical method to estimate nebular oxygen abundance based on the detection of only  strong emission lines calibrated  \citep{Maiolino_2008, Jiang_2019}.  We refer to the latter as the strong-line method.

\subsubsection{Direct Method}\label{sec:direct}
The intensity of  collisionally excited emission lines depends on the ion density, the electron density, and the electron temperature.  The direct method measures the oxygen abundance relative to hydrogen based on the extinction-corrected, emission line ratios [\oiii]$\lambda\lambda$4959, 5007/\hb,  [\oii]$\lambda$3727/\hb, with the electron temperature T$_e$([\oiii])  calculated from the  [\oiii]$\lambda$~5007/[\oiii]$\lambda$~4363 emission line ratio.  Specifically \citep{Izotov_2006}: 
\begin{equation}
\begin{aligned}
    12 & + \text{log}(\frac{\text{O}^{+}}{\text{H}^+}) =  \text{log}\frac{\text{[\oii]$\lambda$3726 +[\oii]$\lambda$3729}}{\text{\hb}} \\ 
       & + 5.961 + \frac{1.672}{t_2} - 0.40 \text{log}(t_2) - 0.034~t_2
\end{aligned}
\end{equation}

\noindent
and
\begin{equation}
\begin{aligned}
    12 & + \text{log}(\frac{\text{O}^{++}}{\text{H}^+}) =  \text{log}\frac{\text{[\oiii]$\lambda$4959 +[\oiii]$\lambda$5007}}{\text{\hb}} \\ 
       & + 6.200 + \frac{1.251}{t_3} - 0.55 \text{log}(t_3) - 0.014~t_3
\end{aligned}
\end{equation}

\noindent
where $t_2 = t_e$(\oii) and $t_3 = t_e$(\oiii) are both in 10$^4$K.  

\subsubsection{Strong Line Calibration}\label{sec:SL}

The direct method is an accurate method to determine the oxygen abundance.  However, often one would have to use the empirical strong line calibration method when the [\oiii]$\lambda$ 4363 is not detected. Strong metal line ratios can be used as indicators, such as [\oiii]$\lambda\lambda$4959,5007/\hb, [\nii]$\lambda\lambda$6548,6583/\ha, [\sii]$\lambda\lambda$6717,6731/\ha, etc. Given the wavelength coverage of our spectra,  we only consider the calibration with the oxygen lines and \hb\ in this work. Here we introduce the commonly used notation for the line ratios: R$_{23}$ = ( [\oiii]$\lambda\lambda$ 4959, 5007 + [\oiii]$\lambda$ 3727)/\hb, and O$_{32}$ = [\oiii]$\lambda$ 5007/[\oii]$\lambda$ 3727. 

\citet{Skillman_1989b}, \citet{McGaugh_1991}, \citet{Pilyugin_2000}, and \citet{van_Zee_2006} have demonstrated that the strong oxygen lines are dependent on both the oxygen abundance and ionization parameters in low metallicity \hii\ regions.
The O$_{32}$ ratio is highly sensitive to the ionization parameter, so it can be used in combination with other ratios to break the degeneracy. \citet{Jiang_2019}  provide a calibration with a combination of R$_{23}$ and O$_{32}$ based on 835 star forming green-pea galaxies, with metallicities ranging from 7.2 $\leq$ 12+log(O/H) $\leq$ 8.6, and 15 galaxies with metallicities lower than 7.6.  Since the metallicities of the BCDs fall into this range, we compare our results of the direct method with this strong line calibration.

The metallicity $Z$ = 12+log(O/H) in this method is determined by solving the equation: 
\begin{equation}
\label{eqn:SL_R23}
\text{log(R23)} = a + b \times Z + c \times Z^2 - d \times (e + Z) \times y, 
\end{equation}
where $y$=log ([\oiii]$\lambda\lambda$4959, 5007/[\oii]$\lambda$3727), which can be approximate as  $y$ =log(O$_{32}$)+0.125, given the intrinsic doublet line ratio [\oiii]$\lambda$ 5007/[\oiii]$\lambda$ 4959 = 2.98 \citep{Storey_2000}.  The coefficients are : $a$ = -24.135, $b$ = 6.1532, $c$ = -0.37866, $d$ = -0.147, $e$ = -7.071. 

To derive 12+log(O/H) as a function of R23 and $y$ from equation~(\ref{eqn:SL_R23}), \citet{Jiang_2019} suggest a different formula based on the values of $y$ and log(R23). However, we found the result more reasonable if we only apply the following formula regardless of the value $y$: 
\begin{equation}
\label{eqn:SL}
\begin{aligned}
\text{12}+&\text{log(O/H)} = \frac{(d \times y - b)}{2c}\\ 
+& \frac{\sqrt{(b-d\times y)^2 -4c\times (a-d\times e\times y - \text{log(R23)})} }{2c}.
\end{aligned}
\end{equation}
We present the results and a comparison in Section~\ref{sec:discuss_SL}.


\subsection{Stellar Mass and Luminosity}\label{sec:mass}
We estimate the stellar mass of the galaxies using the photometric SED fitting code \textsc{FAST++} \citep{Kriek_2009} {\footnote{https://github.com/cschreib/fastpp}}.   
The input data are the emission line corrected galaxy magnitudes  in the $FUV$, $NUV$, $g$, $r$, $i$, and $z$ bands.  
We calculate the emission line correction for the $g$-band magnitude using the available spectra. Specifically, we create a new spectrum where the strongest lines in the wavelength range covered by the filter are masked. We call  f$^{\prime}(\lambda)$ the masked spectrum, and  f$(\lambda)$  the observed spectrum.  We then compute the  emission line corrected  $g$ band magnitude from the observed magnitude, ${g_0}$, as:

\begin{equation}
    g = {g_0} + 2.5\times \text{log}(\frac{\int f(\lambda) P(\lambda) d\lambda }{\int f^{\prime}(\lambda) P(\lambda) d\lambda }), 
\end{equation}

where P($\lambda$) is the throughput of the filter. For the $r$ band we proceed in a different way, because the spectra do not always cover the \ha\ wavelength range. Accordingly,
we calculate the emission-line-free $r$ band magnitude  as follows: 
\begin{equation}
    r = -2.5\times \text{log}( f_{r_0} - \frac{\text{F(\ha)}}{W_r} ) + 48.6, 
\end{equation}
where f$_{r_0}$ is the flux density computed from the $r$ band magnitude before emission-line correction, F(\ha) is the  \ha\ emission line flux estimated from the intrinsic \hb\ flux and the observed dust attenuation, and W$_r$ is the width of the $r$ band filter.  Finally, the GALEX $FUV$ filter covers from 1340 \AA\ to 1800 \AA.  Since all of the objects in our discussion have redshift z $<$ 0.1,  the $FUV$ filter does not cover the Ly$\alpha$ emission, and we do not apply emission line correction on $FUV$. 

In the SED fitting, we used the Flexible Stellar Population Synthesis (FSPS) model \citep{Conroy_2010}, assuming a Chabrier initial mass function (IMF) \citep{Chabrier_2003} and metallicity in the range of Z=0.0008 and Z=0.0031,  corresponding  to 5$\%$ to 20$\%$ solar metallicity.  We describe the star formation history with a ``tau model" (SFR($t$) $\propto$ $exp(-t/\tau)$), where $t$ is the age of the galaxy and  $\tau$ varies from 6.5 Myr to 10 Myr in steps of 0.5 Myr. 

\citet{Hsyu_2018} estimate the stellar masses of the galaxies in their sample using a color-dependent mass-to-light ratio provided in \citet{Bell_2003}.  However, this relation was derived from objects with masses larger than 10$^{8.5}$ M$_\odot$, not optimized for low mass objects with starbursts, like those studied in this paper and in  Hsyu et al (2018).  For consistency, we calculated the stellar masses of BCDs in \citet{Hsyu_2018} with the same SED fitting analysis described here. 

Finally, to compare with other results in the literature, we need to derive the luminosity of the galaxies in the $B-$band filter. We used the empirical relation between the $B$ band and the $i-$band and $g-i$ color, given in \citet{Cook_2014}: 

\begin{equation}
\text{B} = i + (1.27 \pm 0.03)\times(g - i)+ (0.16\pm 0.01),
\end{equation}

where we used the $g$ emission-line-free magnitude.

\subsection{Star Formation Rate}\label{sec:SFR}
We calculate the star formation rate with both the \hb\ luminosity and UV continuum luminosity L$_\nu$(1500) using the method in \citet{K98}. Throughout the paper, we denote the SFR derived from \hb\ flux as SFR and SFR derived from UV continuum as SFR(UV). The specific star formation rate (sSFR) is defined as sSFR = SFR/M$_\ast$.  
The reported \hb\ luminosities, L(\hb), are calculated using the slit-loss corrected \hb\ line flux. We estimate the slit-loss corrected flux by assuming a Gaussian profile with the width similar to the seeing, and calculate the slit loss for the assumed profile observed with a 0.9\arcsec-wide long slit. 
We estimate the SFR with the extinction-corrected \hb\ luminosity adopting an intrinsic \ha/\hb\ ratio of 2.86: 

\begin{equation}
    \text{SFR} = \epsilon(\text{IMF})\times 2.21 \times10^{-41} \text{L(\hb)},
\end{equation}
where  L(\hb) is in  units of erg s$^{-1}$ and the SFR is in M$_\odot$ yr$^{-1}$.  $\epsilon(\text{IMF})$ is the factor to correct the flattening of the stellar IMF below 1 M$_\odot$ for a Chabrier IMF, instead of the power law Salpeter IMF \citep{Salpeter_1955} adopted by \citet{K98}: 

\begin{equation}
    \epsilon(\text{IMF}) = \frac{\int^{100}_{0.1} m\times\xi_{Ch}(m) dm }{\int^{100}_{0.1} m\times\xi_{Sal}(m) dm} \simeq 1/1.46, 
\end{equation}
where $\xi_{Ch}(m)$ and $\xi_{Sal}(m)$ are the Chabrier IMF and Salpeter IMF, respectively.  The SFR of our samples ranges from 0.03 to 3 M$_\odot$ yr$^{-1}$. At these low levels of SFR one would normally have to worry about the sampling of the high mass end of the IMF. However, the presence of strong [\oiii] emission lines shows that the most massive stars are still alive in the current bursts.  

We calculate the UV continuum L$_\nu$(1500\AA) from the flux density of GALEX $FUV$ magnitudes, and correct the dust extinction. 
The SFR(UV) is calculated as \citep{McQuinn_2015}: 
\begin{equation}\label{eq:SFR_UV}
    \text{SFR(UV)} = \epsilon(\text{IMF}) \times (2.04 \pm 0.81) \times 10^{-28} L_\nu(1500\text{\AA}). 
\end{equation}
The SFR(UV) of our samples ranges from 0.02 to 10 M$_\odot$ yr$^{-1}$. 
The uncertainties of the SFR(UV) are computed by generating 500 realizations by randomizing the E(B-V) and $FUV$ magnitudes within $\pm 1\sigma$, and combined with the uncertainty in equation~\ref{eq:SFR_UV}.  For each realization, we compute the SFR(UV). We report the 50$\%$ percentiles as the SFR(UV), and 16$\%$ and 84$\%$ percentiles of the distribution.



\begin{table*} 
\centering 
\caption{The extinction corrected lines fluxes of the observed BCDs, normalized to \hb. The \hb\ flux is reported in the unit of 10$^{-15}$ erg s$^{-1}$ cm$^{-2}$.  The 3$\sigma$ upper limit is reported for low S/N emission lines. } 
\begin{threeparttable} 
\begin{tabular*}{1.05\textwidth}{cccccccc}   
 ID & [\oii ]$_{tot}$/\hb  & \hd/\hb & \hr/\hb & [\oiii ]$\lambda$ 4363/\hb & \hb & [\oiii ]$\lambda$ 4959/\hb & [\oiii ]$\lambda$ 5007/\hb \\ 
  &   &  &  &  & [ 10$^{-15}$ erg/s/cm$^{2}$ ] &  &  \\ 
 \hline \hline 
J0006-5911 & 0.59$\pm$0.035 & 0.21$\pm$0.02 & 0.48$\pm$0.02 & 0.05$\pm$0.01 & 6.18$\pm$0.16 & 0.87$\pm$0.02 & 2.60$\pm$0.06 \\ 
J0019-5937 & 1.35$\pm$0.04 & 0.27$\pm$0.07 & 0.48$\pm$0.05 & 0.08$\pm$0.01 & 4.25$\pm$0.19 & 1.74$\pm$0.04 & 5.35$\pm$0.11 \\ 
J0212-3231 & 0.54$\pm$0.03 & 0.27$\pm$0.05 & 0.48$\pm$0.03 & 0.10$\pm$0.01 & 6.94$\pm$0.28 & 1.99$\pm$0.04 & 5.98$\pm$0.12 \\ 
J0218-1946 & 0.85$\pm$0.03 & 0.27$\pm$0.05 & 0.50$\pm$0.03 & 0.10$\pm$0.01 & 2.69$\pm$0.10 & 1.83$\pm$0.04 & 5.45$\pm$0.12 \\ 
J0218-0912 & 0.78$\pm$0.02 & 0.27$\pm$0.01 & 0.46$\pm$0.01 & 0.14$\pm$0.01 & 21.51$\pm$0.44 & 2.35$\pm$0.05 & 7.01$\pm$0.14 \\ 
J0242-5149 & $<$0.07 & $<$0.38 & 0.49$\pm$0.07 & 0.08$\pm$0.01 & 0.81$\pm$0.05 & 1.29$\pm$0.03 & 3.86$\pm$0.08 \\ 
J0249-0111 & 0.89$\pm$0.04 & 0.26$\pm$0.02 & 0.47$\pm$0.02 & 0.07$\pm$0.01 & 11.42$\pm$0.28 & 1.80$\pm$0.04 & 5.36$\pm$0.12 \\ 
J0311-1341 & 0.86$\pm$0.03 & 0.27$\pm$0.03 & 0.48$\pm$0.02 & 0.08$\pm$0.01 & 2.63$\pm$0.10 & 1.35$\pm$0.03 & 4.09$\pm$0.08 \\ 
J0426-5410 & 1.78$\pm$0.05 & 0.27$\pm$0.07 & 0.48$\pm$0.06 & 0.04$\pm$0.01 & 6.37$\pm$0.31 & 1.29$\pm$0.03 & 3.87$\pm$0.08 \\ 
J0450-5429 & 1.32$\pm$0.04 & 0.27$\pm$0.06 & 0.48$\pm$0.04 & 0.05$\pm$0.01 & 2.71$\pm$0.11 & 1.12$\pm$0.03 & 3.35$\pm$0.07 \\ 
J0500-3204 & 1.30$\pm$0.04 & 0.27$\pm$0.02 & 0.47$\pm$0.02 & 0.10$\pm$0.01 & 3.42$\pm$0.10 & 1.59$\pm$0.03 & 4.76$\pm$0.10 \\ 
J0520-5111 & 1.41$\pm$0.05 & 0.26$\pm$0.01 & 0.47$\pm$0.01 & 0.10$\pm$0.01 & 13.91$\pm$0.31 & 1.72$\pm$0.04 & 5.35$\pm$0.11 \\ 
J0522-2251 & 0.25$\pm$0.02 & 0.27$\pm$0.01 & 0.46$\pm$0.01 & 0.10$\pm$0.01 & 3.80$\pm$0.08 & 1.31$\pm$0.03 & 3.93$\pm$0.08 \\ 
J0524-2041 & 0.44$\pm$0.02 & 0.27$\pm$0.01 & 0.46$\pm$0.01 & 0.15$\pm$0.01 & 11.92$\pm$0.25 & 1.69$\pm$0.03 & 5.20$\pm$0.11 \\ 
J0527-2704 & 0.70$\pm$0.02 & 0.27$\pm$0.02 & 0.48$\pm$0.02 & 0.14$\pm$0.01 & 6.96$\pm$0.17 & 2.00$\pm$0.04 & 5.94$\pm$0.12 \\ 
J0617-4954 & 0.63$\pm$0.04 & 0.27$\pm$0.02 & 0.48$\pm$0.02 & 0.14$\pm$0.01 & 9.86$\pm$0.23 & 1.36$\pm$0.03 & 4.06$\pm$0.09 \\ 
J2014-5348 & 1.07$\pm$0.04 & 0.27$\pm$0.04 & 0.48$\pm$0.03 & 0.17$\pm$0.01 & 1.44$\pm$0.06 & 1.72$\pm$0.04 & 5.48$\pm$0.12 \\ 
J2030-4719 & 0.92$\pm$0.03 & 0.27$\pm$0.03 & 0.48$\pm$0.04 & 0.08$\pm$0.01 & 2.69$\pm$0.10 & 1.32$\pm$0.03 & 3.93$\pm$0.09 \\ 
J2036-4918 & 1.28$\pm$0.05 & 0.27$\pm$0.01 & 0.44$\pm$0.01 & 0.07$\pm$0.01 & 2.86$\pm$0.07 & 1.11$\pm$0.03 & 3.30$\pm$0.08 \\ 
J2044-5731 & 0.78$\pm$0.05 & 0.22$\pm$0.02 & 0.41$\pm$0.02 & 0.10$\pm$0.01 & 4.23$\pm$0.12 & 1.05$\pm$0.025 & 3.136$\pm$0.066 \\ 
J2051-4559 & 1.08$\pm$0.04 & 0.27$\pm$0.05 & 0.49$\pm$0.04 & 0.08$\pm$0.01 & 1.59$\pm$0.06 & 1.00$\pm$0.03 & 3.01$\pm$0.07 \\ 
J2052-4517 & 1.18$\pm$0.05 & 0.26$\pm$0.02 & 0.47$\pm$0.02 & 0.09$\pm$0.01 & 3.74$\pm$0.11 & 1.77$\pm$0.04 & 5.28$\pm$0.12 \\ 
J2055-5405 & 1.07$\pm$0.04 & 0.27$\pm$0.01 & 0.46$\pm$0.02 & 0.09$\pm$0.01 & 2.83$\pm$0.07 & 1.43$\pm$0.03 & 4.27$\pm$0.10 \\ 
J2109-5228 & 1.52$\pm$0.04 & 0.27$\pm$0.06 & 0.48$\pm$0.05 & 0.10$\pm$0.01 & 3.95$\pm$0.18 & 1.56$\pm$0.04 & 4.64$\pm$0.10 \\ 
J2117-5920 & 1.17$\pm$0.04 & 0.24$\pm$0.02 & 0.40$\pm$0.02 & 0.07$\pm$0.01 & 3.56$\pm$0.10 & 1.52$\pm$0.03 & 4.53$\pm$0.10 \\ 
J2120-5054 & 1.68$\pm$0.07 & 0.26$\pm$0.02 & 0.45$\pm$0.02 & $<$0.04 & 7.63$\pm$0.21 & 1.08$\pm$0.03 & 3.34$\pm$0.07 \\ 
J2203-4013 & 1.06$\pm$0.03 & 0.27$\pm$0.01 & 0.45$\pm$0.01 & 0.08$\pm$0.01 & 4.39$\pm$0.10 & 1.39$\pm$0.03 & 4.26$\pm$0.09 \\ 
J2205-6303 & 1.20$\pm$0.04 & 0.28$\pm$0.01 & 0.47$\pm$0.01 & 0.08$\pm$0.01 & 5.09$\pm$0.12 & 1.91$\pm$0.04 & 5.74$\pm$0.12 \\ 
J2205-4218 & 1.26$\pm$0.05 & 0.27$\pm$0.04 & 0.48$\pm$0.03 & 0.09$\pm$0.01 & 6.86$\pm$0.22 & 1.29$\pm$0.03 & 3.85$\pm$0.09 \\ 
J2212-4527 & 1.35$\pm$0.04 & 0.26$\pm$0.01 & 0.47$\pm$0.01 & 0.06$\pm$0.01 & 4.42$\pm$0.10 & 1.63$\pm$0.04 & 4.87$\pm$0.10 \\ 
J2213-4320 & 1.83$\pm$0.05 & 0.26$\pm$0.03 & 0.47$\pm$0.02 & 0.08$\pm$0.01 & 4.47$\pm$0.13 & 1.51$\pm$0.03 & 4.59$\pm$0.10 \\ 
J2237-4845 & 2.31$\pm$0.07 & 0.26$\pm$0.02 & 0.47$\pm$0.02 & 0.09$\pm$0.01 & 8.53$\pm$0.22 & 1.47$\pm$0.03 & 4.41$\pm$0.09 \\ 
J1403+0151 & 2.28$\pm$0.17 & 0.28$\pm$0.03 & 0.38$\pm$0.03 & 0.12$\pm$0.02 & 17.46$\pm$0.48 & 0.92$\pm$0.03 & 2.82$\pm$0.06 \\ 
J1409+0116 & 1.77$\pm$0.06 & 0.27$\pm$0.05 & 0.48$\pm$0.04 & 0.06$\pm$0.01 & 1.29$\pm$0.06 & 0.83$\pm$0.02 & 2.52$\pm$0.06 \\ 
J1451+0231 & 2.32$\pm$0.08 & 0.21$\pm$0.02 & 0.40$\pm$0.02 & 0.02$\pm$0.01 & 13.06$\pm$0.34 & 1.07$\pm$0.02 & 3.28$\pm$0.07 \\ 
\\
\hline 
\end{tabular*} 
\label{tab:emission} 
\end{threeparttable} 
\end{table*}

\begin{table*} 
\centering 
\caption{The derived properties for the observed BCDs.  For object J2120-5054, the [\oiii]$\lambda$4363 has S/N $<$ 5 ,therefore the T$_e$[\oiii] are not reported. } 
\begin{threeparttable} 
\begin{tabular*}{0.98\textwidth}{ccccccccc} 
 ID & redshift & T$_e$ & 12+log(O/H) & SFR  & SFR(UV)  & Log(M$_{*}$/M$_\odot$) & E(B-V) & Mag B \\
   &  &  [K] &  &  [M$_\odot$ yr$^{-1}$] &  [M$_\odot$ yr$^{-1}$] &  &  &  \\
\hline\hline 
 J0006-5911 & 0.0109 &  15200$^{+100}_{-100}$ &7.55$^{+0.01}_{-0.01}$ &0.040$^{+0.002}_{-0.002}$ &0.012$^{+0.007}_{-0.006}$ &6.73$^{+0.34}_{-0.01}$ &0.15$^{+0.02}_{-0.02}$ &-13.53$\pm$0.05\\ 
 J0019-5937 & 0.0357 &  13700$^{+100}_{-100}$ &7.97$^{+0.01}_{-0.01}$ &0.309$^{+0.077}_{-0.059}$ &0.141$^{+0.069}_{-0.064}$ &6.26$^{+0.12}_{-0.10}$ &0.00$^{+0.04}_{-0.03}$ &-17.14$\pm$0.04\\ 
 J0212-3231 & 0.0490 &  14200$^{+100}_{-200}$ &7.92$^{+0.01}_{-0.01}$ &0.802$^{+0.218}_{-0.228}$ &9.600$^{+7.100}_{-5.076}$ &7.81$^{+0.06}_{-0.01}$ &0.49$^{+0.06}_{-0.05}$ &-16.48$\pm$0.04\\ 
 J0218-1946 & 0.0304 &  14900$^{+100}_{-100}$ &7.85$^{+0.01}_{-0.01}$ &0.121$^{+0.018}_{-0.018}$ &0.122$^{+0.063}_{-0.057}$ &7.17$^{+0.03}_{-0.02}$ &0.08$^{+0.03}_{-0.03}$ &-14.99$\pm$0.05\\ 
 J0218-0912 & 0.0127 &  15200$^{+100}_{-100}$ &7.93$^{+0.01}_{-0.01}$ &0.126$^{+0.016}_{-0.017}$ &0.155$^{+0.089}_{-0.074}$ &8.25$^{+0.05}_{-0.12}$ &0.41$^{+0.03}_{-0.02}$ &-14.65$\pm$0.04\\ 
 J0242-5149 & 0.0526 &  15400$^{+200}_{-100}$ &7.60$^{+0.01}_{-0.01}$ &0.123$^{+0.032}_{-0.024}$ &0.502$^{+0.352}_{-0.262}$ &8.12$^{+0.15}_{-0.04}$ &0.00$^{+0.07}_{-0.04}$ &-16.11$\pm$0.05\\ 
 J0249-0111 & 0.0672 &  13100$^{+400}_{-400}$ &7.99$^{+0.03}_{-0.03}$ &2.713$^{+1.525}_{-1.617}$ &12.021$^{+22.690}_{-8.930}$ &7.25$^{+0.01}_{-0.24}$ &0.57$^{+0.11}_{-0.13}$ &-17.98$\pm$0.04\\ 
 J0311-1341 & 0.0311 &  15100$^{+200}_{-200}$ &7.73$^{+0.01}_{-0.01}$ &0.124$^{+0.034}_{-0.033}$ &0.388$^{+0.390}_{-0.230}$ &7.51$^{+0.01}_{-0.10}$ &0.17$^{+0.08}_{-0.08}$ &-15.06$\pm$0.05\\ 
 J0426-5410 & 0.0551 &  11600$^{+100}_{-100}$ &8.08$^{+0.01}_{-0.01}$ &0.929$^{+0.159}_{-0.192}$ &1.655$^{+0.952}_{-0.837}$ &7.13$^{+0.26}_{-0.03}$ &0.30$^{+0.04}_{-0.05}$ &-17.14$\pm$0.04\\ 
 J0450-5429 & 0.0471 &  14000$^{+200}_{-300}$ &7.79$^{+0.01}_{-0.01}$ &0.279$^{+0.098}_{-0.106}$ &-- &7.82$^{+0.22}_{-0.02}$ &0.21$^{+0.13}_{-0.12}$ &-17.17$\pm$0.04\\ 
 J0500-3204 & 0.0401 &  15900$^{+300}_{-300}$ &7.78$^{+0.01}_{-0.01}$ &0.249$^{+0.073}_{-0.094}$ &0.286$^{+0.461}_{-0.204}$ &7.51$^{+0.02}_{-0.58}$ &0.18$^{+0.12}_{-0.12}$ &-16.00$\pm$0.05\\ 
 J0520-5111 & 0.0569 &  15000$^{+300}_{-600}$ &7.88$^{+0.02}_{-0.02}$ &2.225$^{+0.755}_{-1.094}$ &-- &8.34$^{+0.11}_{-0.14}$ &0.59$^{+0.09}_{-0.11}$ &-16.56$\pm$0.04\\ 
 J0522-2251 & 0.0310 &  17300$^{+200}_{-100}$ &7.52$^{+0.01}_{-0.01}$ &0.196$^{+0.033}_{-0.034}$ &0.673$^{+0.338}_{-0.320}$ &8.14$^{+0.04}_{-0.03}$ &0.12$^{+0.04}_{-0.04}$ &-14.79$\pm$0.07\\ 
 J0524-2041 & 0.0093 &  18600$^{+200}_{-100}$ &7.59$^{+0.01}_{-0.01}$ &0.038$^{+0.006}_{-0.005}$ &-- &8.17$^{+0.03}_{-0.25}$ &0.35$^{+0.03}_{-0.03}$ &-12.52$\pm$0.05\\ 
 J0527-2704 & 0.0534 &  16900$^{+600}_{-700}$ &7.73$^{+0.03}_{-0.03}$ &1.093$^{+0.588}_{-0.577}$ &1.444$^{+2.684}_{-1.070}$ &7.67$^{+0.21}_{-0.18}$ &0.34$^{+0.12}_{-0.13}$ &-16.89$\pm$0.04\\ 
 J0617-4954 & 0.0512 &  20500$^{+500}_{-500}$ &7.43$^{+0.02}_{-0.02}$ &1.238$^{+0.377}_{-0.399}$ &13.014$^{+16.525}_{-8.908}$ &6.73$^{+0.13}_{-0.01}$ &0.57$^{+0.09}_{-0.11}$ &-15.88$\pm$0.05\\ 
 J2014-5348 & 0.0584 &  16600$^{+500}_{-200}$ &7.76$^{+0.01}_{-0.01}$ &0.334$^{+0.178}_{-0.098}$ &0.212$^{+0.184}_{-0.123}$ &5.91$^{+0.63}_{-0.01}$ &0.01$^{+0.09}_{-0.05}$ &-16.28$\pm$0.06\\ 
 J2030-4719 & 0.0478 &  15100$^{+400}_{-200}$ &7.73$^{+0.01}_{-0.01}$ &0.412$^{+0.255}_{-0.141}$ &-- &7.77$^{+0.25}_{-0.04}$ &0.00$^{+0.14}_{-0.07}$ &-16.68$\pm$0.04\\ 
 J2036-4918 & 0.0251 &  15800$^{+100}_{-100}$ &7.67$^{+0.01}_{-0.01}$ &0.105$^{+0.017}_{-0.016}$ &0.094$^{+0.045}_{-0.043}$ &7.33$^{+0.04}_{-0.03}$ &0.16$^{+0.04}_{-0.03}$ &-15.08$\pm$0.05\\ 
 J2044-5731 & 0.0249 &  20100$^{+100}_{-200}$ &7.39$^{+0.01}_{-0.01}$ &0.177$^{+0.006}_{-0.013}$ &-- &7.25$^{+0.12}_{-0.09}$ &0.09$^{+0.02}_{-0.04}$ &-15.42$\pm$0.04\\ 
 J2051-4559 & 0.0499 &  17300$^{+100}_{-100}$ &7.54$^{+0.01}_{-0.01}$ &0.313$^{+0.041}_{-0.043}$ &0.715$^{+0.360}_{-0.339}$ &7.61$^{+0.18}_{-0.11}$ &0.00$^{+0.05}_{-0.03}$ &-16.76$\pm$0.05\\ 
 J2052-4517 & 0.0309 &  14300$^{+200}_{-200}$ &7.91$^{+0.01}_{-0.01}$ &0.248$^{+0.092}_{-0.083}$ &0.096$^{+0.089}_{-0.056}$ &7.38$^{+-0.01}_{-0.45}$ &0.06$^{+0.07}_{-0.06}$ &-16.02$\pm$0.04\\ 
 J2055-5405 & 0.0622 &  15500$^{+200}_{-200}$ &7.74$^{+0.01}_{-0.01}$ &0.727$^{+0.146}_{-0.177}$ &0.391$^{+0.314}_{-0.211}$ &7.25$^{+0.19}_{-0.04}$ &0.11$^{+0.04}_{-0.07}$ &-17.12$\pm$0.04\\ 
 J2109-5228 & 0.0498 &  15800$^{+300}_{-100}$ &7.80$^{+0.01}_{-0.01}$ &0.579$^{+0.211}_{-0.135}$ &-- &7.40$^{+0.19}_{-0.18}$ &0.00$^{+0.08}_{-0.05}$ &-17.56$\pm$0.04\\ 
 J2117-5920 & 0.0582 &  13800$^{+100}_{-100}$ &7.90$^{+0.01}_{-0.01}$ &0.796$^{+0.027}_{-0.059}$ &0.281$^{+0.164}_{-0.136}$ &7.61$^{+0.08}_{-0.03}$ &0.10$^{+0.02}_{-0.04}$ &-16.94$\pm$0.05\\ 
 J2120-5054 & 0.0434 &  -- &7.60$^{+0.01}_{-0.01}$ &1.019$^{+0.264}_{-0.204}$ &8.965$^{+5.371}_{-4.395}$ &7.68$^{+0.04}_{-0.24}$ &0.46$^{+0.05}_{-0.04}$ &-17.04$\pm$0.04\\ 
 J2203-4013 & 0.0366 &  15000$^{+100}_{-100}$ &7.78$^{+0.01}_{-0.01}$ &0.311$^{+0.045}_{-0.044}$ &-- &8.23$^{+0.09}_{-0.04}$ &0.14$^{+0.03}_{-0.03}$ &-15.92$\pm$0.05\\ 
 J2205-6303 & 0.0542 &  13400$^{+100}_{-100}$ &8.01$^{+0.01}_{-0.01}$ &0.793$^{+0.000}_{-0.000}$ &0.106$^{+0.057}_{-0.052}$ &7.76$^{+0.05}_{-0.27}$ &0.07$^{+0.01}_{-0.01}$ &-16.77$\pm$0.04\\ 
 J2205-4218 & 0.0270 &  16900$^{+1000}_{-900}$ &7.65$^{+0.02}_{-0.02}$ &0.323$^{+0.273}_{-0.241}$ &1.229$^{+3.607}_{-1.045}$ &7.89$^{+0.03}_{-0.26}$ &0.40$^{+0.18}_{-0.17}$ &-15.42$\pm$0.04\\ 
 J2212-4527 & 0.0656 &  12900$^{+200}_{-400}$ &8.00$^{+0.02}_{-0.01}$ &1.063$^{+0.303}_{-0.506}$ &2.891$^{+4.191}_{-2.014}$ &7.48$^{+0.07}_{-0.04}$ &0.23$^{+0.07}_{-0.15}$ &-16.80$\pm$0.05\\ 
 J2213-4320 & 0.0679 &  14100$^{+300}_{-400}$ &7.93$^{+0.01}_{-0.01}$ &1.129$^{+0.435}_{-0.530}$ &3.249$^{+6.926}_{-2.427}$ &7.62$^{+0.13}_{-0.01}$ &0.24$^{+0.13}_{-0.15}$ &-17.30$\pm$0.04\\ 
 J2237-4845 & 0.0502 &  15900$^{+400}_{-600}$ &7.84$^{+0.01}_{-0.01}$ &1.150$^{+0.417}_{-0.527}$ &3.003$^{+2.614}_{-1.881}$ &7.23$^{+0.02}_{-0.24}$ &0.35$^{+0.07}_{-0.08}$ &-17.91$\pm$0.04\\ 
 J1403+0151 & 0.0243 &  21900$^{+100}_{-100}$ &7.45$^{+0.01}_{-0.01}$ &0.705$^{+0.138}_{-0.147}$ &77.625$^{+43.361}_{-37.433}$ &7.43$^{+0.24}_{-0.02}$ &0.95$^{+0.04}_{-0.04}$ &-15.45$\pm$0.24\\ 
 J1409+0116 & 0.0254 &  16500$^{+400}_{-200}$ &7.61$^{+0.01}_{-0.01}$ &0.053$^{+0.023}_{-0.017}$ &0.157$^{+0.135}_{-0.097}$ &7.72$^{+0.16}_{-0.26}$ &0.04$^{+0.10}_{-0.06}$ &-16.45$\pm$0.08\\ 
 J1451+0231 & 0.0068 &  10500$^{+100}_{-100}$ &8.19$^{+0.01}_{-0.01}$ &0.037$^{+0.002}_{-0.002}$ &0.010$^{+0.005}_{-0.004}$ &8.24$^{+0.09}_{-0.60}$ &0.15$^{+0.02}_{-0.02}$ &-13.90$\pm$0.06\\ 
\hline 
\end{tabular*} 
\label{tab:property} 
\end{threeparttable} 
\end{table*}


\section{Discussion} \label{sec:discussion}

In this section, we compare the galaxy properties of the selected BCDs in the DES with samples of similar galaxies from the literature. Specifically, we consider the sub-sample of BCDs in \citet{Hsyu_2018} for which the metallicity is measured via the direct method, the LVL galaxies in \citet{Berg_2012} with their SFR reported in \citet{Lee_2009}, and the extremely-metal poor (XMP) galaxies with 12+log(O/H) $\leq$ 7.35 (1/20 Z$_\odot$), defined in \citet{McQuinn_2020}.  The galaxies in these samples are all at low redshift ($<$0.1) with 12+log(O/H)$<8.25$ (0.4 Z$_\odot$) determined with the direct method.

\subsection{Specific Star Formation Rate}

\begin{figure}
\includegraphics[width=0.99\linewidth]{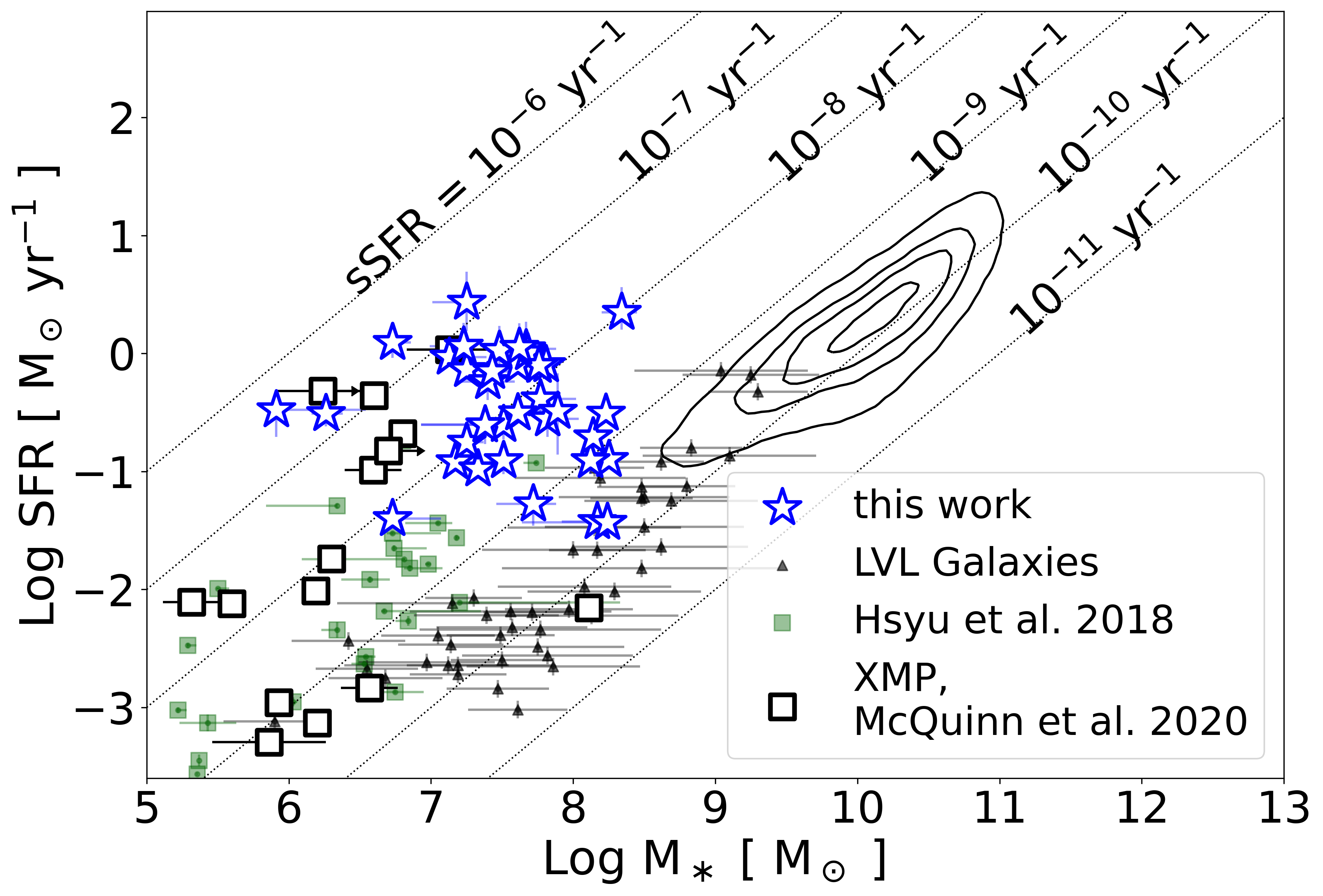}
\caption{ The stellar mass vs. Log(SFR(\hb)). Blue stars are the DES BCDs observed in the work, the green squares are the BCDs reported in \citet{Hsyu_2018}, the black triangles are the LVL galaxies in \citet{Berg_2012}, and the open squares are the XMP galaxies in \citet{McQuinn_2020}. The black contours are typical SFG derived from the SDSS DR8 catalog. sSFRs are indicated with the dotted lines.  \label{fig:sSFR}}
\end{figure}

Figure~\ref{fig:sSFR} shows the SFR(\hb) and stellar mass of the DES BCDs (blue stars), those identified in \citet[][green squares]{Hsyu_2018}, as well as the LVL galaxies from \citet[][black dots]{Lee_2009} and the XMP galaxies in \citet{McQuinn_2020}. For reference, we also show the distribution of star forming galaxies (SFG) with higher masses from the SDSS DR8 \citep{SDSS_DR8} at redshift $\sim$ 0. 
The LVL galaxies extend to lower masses the ``main sequence" of star forming galaxies defined by the SDSS galaxies, with sSFR $\simeq$ 10$^{-10}$ yr$^{-1}$. 
The BCDs in \citet{Hsyu_2018} have stellar masses $<$ 10$^{8.5}$M$_\odot$, and sSFRs mostly vary from 10$^{-10}$ yr$^{-1}$ to 10$^{-8}$ yr$^{-1}$.  
The DES BCDs also have a wide range of sSFRs, and the SFRs are overall much higher.  

\begin{figure}
\includegraphics[width=0.99\linewidth]{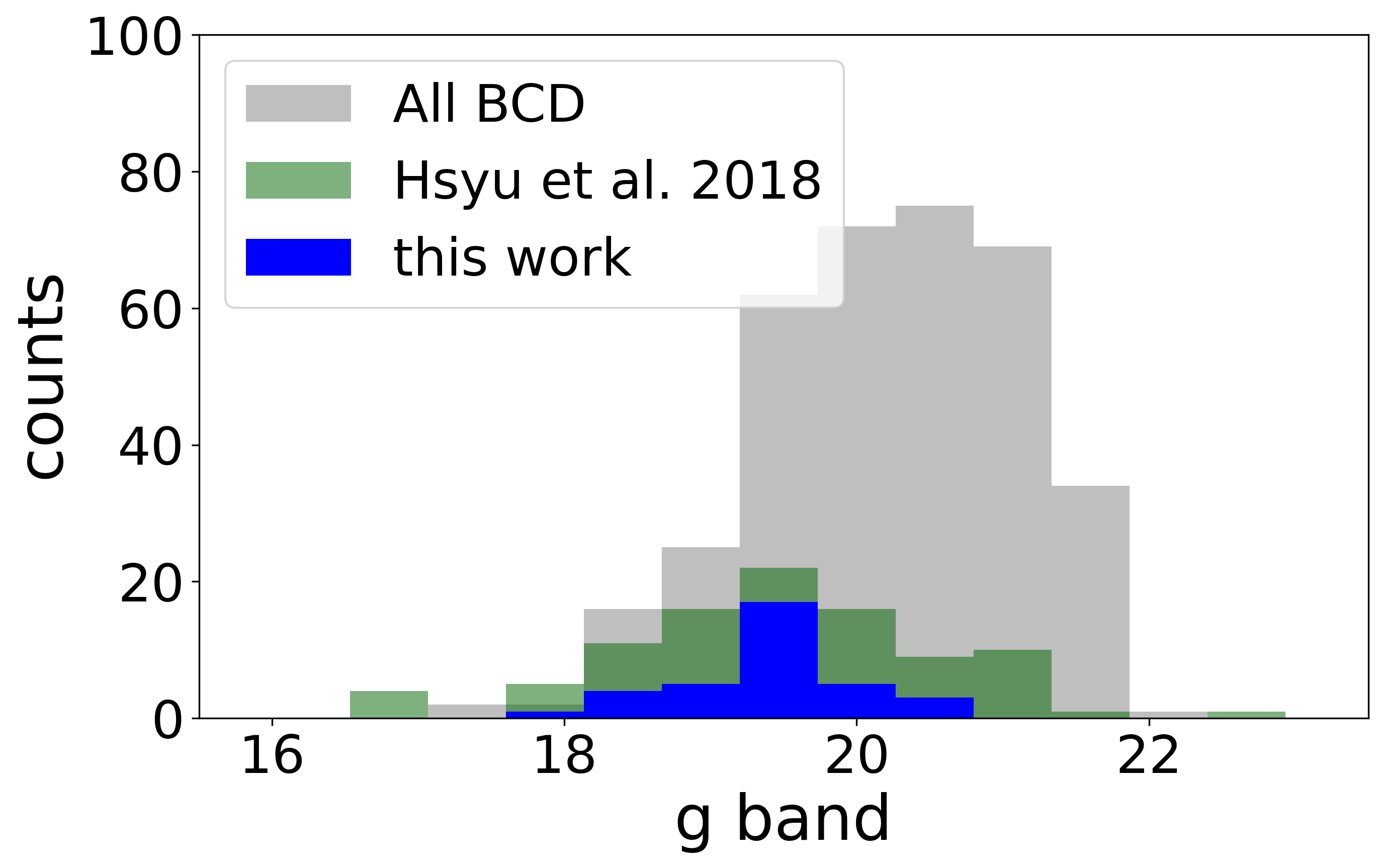}
\caption{ Number counts of $g$-band from the 358 DES BCD candidates (gray), BCDs selected in \citet{Hsyu_2018}, and the 35 BCDs presented in this work.   \label{fig:counts}}
\end{figure}

The star formation rates of the DES BCDs are higher for the following reason: 
{\bf We only obtained long spectroscopic exposures with the b2k disperser if the redshift of the galaxy was higher than 0.02, in order to cover the  [\oii]$\lambda$3727 emission line.} 
We also apply a deeper magnitude cutoff than Hsyu et al. ($g > 22$ and $r > 22$, see Figure~\ref{fig:counts}), the redshift distribution of our galaxy sample is skewed toward slightly higher redshift. In fact the mean redshifts of our BCDs and Hsyu et al.'s sample are 0.041 and 0.015, respectively. Additionally, we required objects to be detected in the GALEX $NUV$ band. For the same $NUV$ magnitudes, higher redshifts imply higher SFRs.

\subsection{Luminosity-Metallicity Relation and Mass-Metallicity Relation}

\begin{figure}[h]
\includegraphics[width=0.99\linewidth]{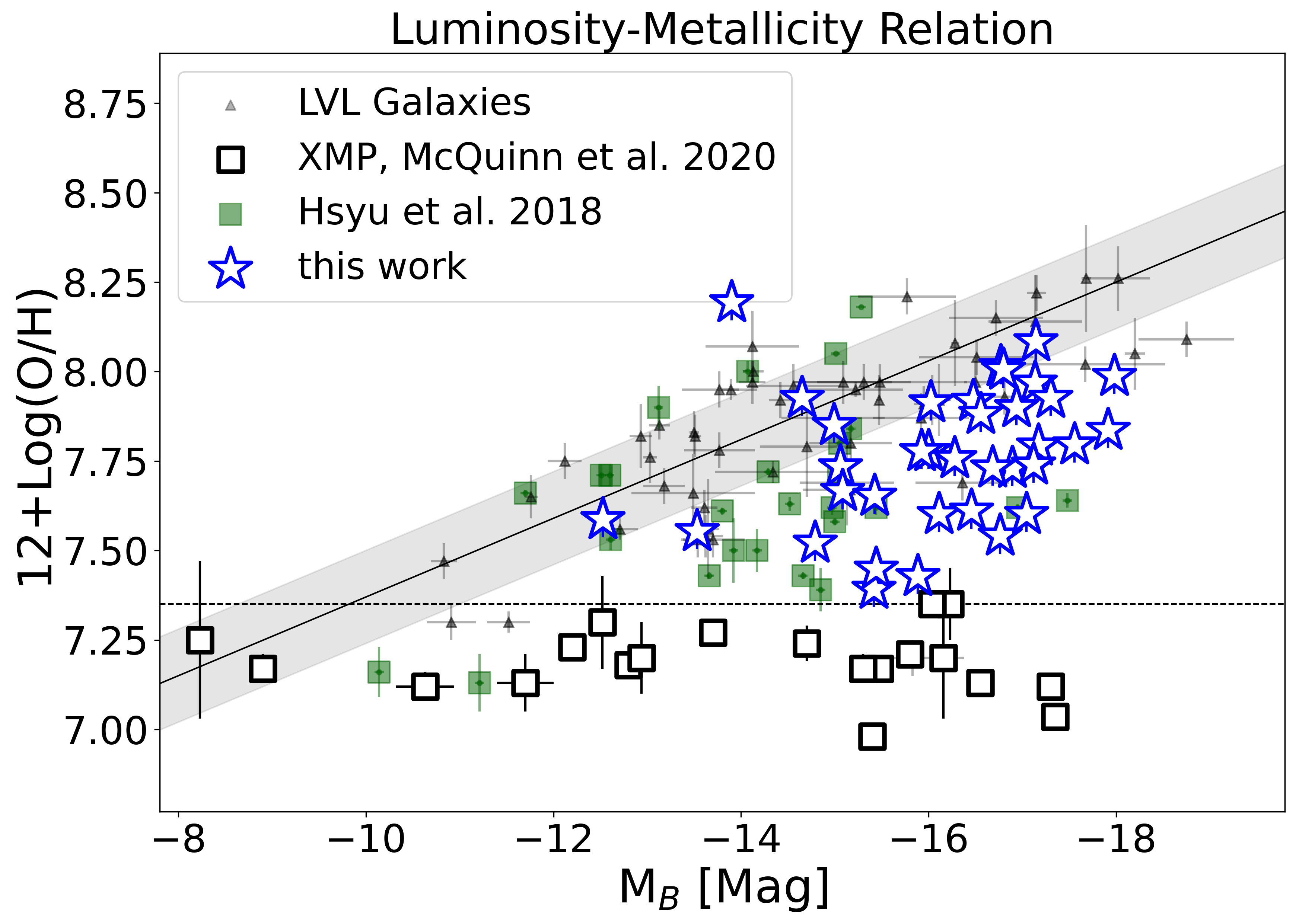}
\caption{ Absolute B band Magnitude vs. Metallicity of our spectroscopic sub-sample (blue stars), BCDs selected in \citet{Hsyu_2018} (green squares), Local Volume Legacy galaxies selected in \citet{Berg_2012} (black triangle dots), and the XMP galaxies \citep{McQuinn_2020} (open squares).  The 1$\sigma$ scatter of the L-Z relation (equation~\ref{eqn:LZ}) is shown in gray shaded areas. \label{fig:LZ}}
\end{figure}

\begin{figure}[h]
\includegraphics[width=0.99\linewidth]{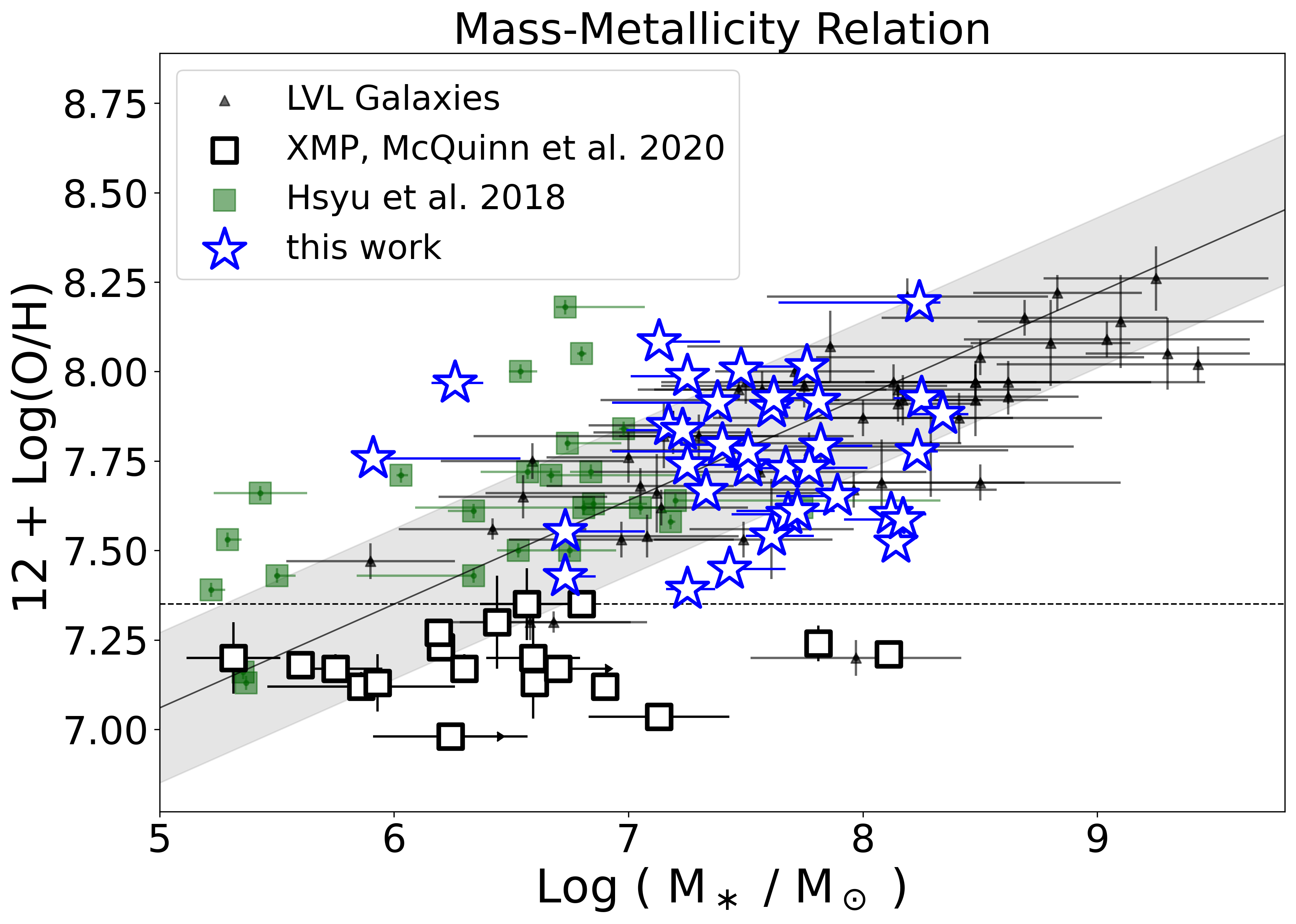}
\caption{ Stellar mass vs. Metallicity of the same galaxy samples as in Figure~\ref{fig:LZ}. The 1$\sigma$ scatter of the M-Z relation (equation~\ref{eqn:MZ}) is shown in gray shaded areas.  Note the tighness of this relationship compared to that in Figure~\ref{fig:LZ}.  \label{fig:MZ}}
\end{figure}

Local Volume Legacy galaxies follow the linear L-Z  and M-Z relation given in \citet{Berg_2012}: 

\begin{equation}
\label{eqn:LZ}
12+\text{log(O/H)}  = (-0.11 \pm 0.01)\times \text{M$_B$} +(6.27 \pm 0.21),
\end{equation}
\begin{equation}
\label{eqn:MZ}
12+\text{log(O/H)}  = (0.29\pm 0.03) \times \text{log}(M_\ast) + (5.31 \pm 0.24).
\end{equation}
We show how our galaxies compare with the L-Z and M-Z relations in Figure~\ref{fig:LZ} and Figure~\ref{fig:MZ}, respectively. 

In Figure~\ref{fig:LZ}, most of our BCDs are brighter than the  $B-$band luminosity  predicted from their metallicity and lie in between the LVL galaxies and the XMP galaxies.  This is not surprising considering that the SFRs of the BCDs are biased high. The high SFR indicates that more young, massive, bright stars are currently present in the galaxy. These massive stars contribute a significant luminosity to the $B-$band, making the BCDs brighter than the L-Z relation.
In terms of stellar mass, massive stars are not as impactful as in luminosity. For example, an O-star can be as bright as 10$^5$ L$_\odot$, with only 20 M$_\odot$. 
Therefore, the M-Z relation is not strongly biased by  recent star formation and shows a tighter relation among the low mass galaxies.
As shown in Figure~\ref{fig:MZ}, most of the BCDs are distributed within 1$\sigma$ from the M-Z relation derived by the LVL galaxies, and many of the XMP galaxies that were outliers in the L-Z relation are also now close to the M-Z trend line. 

Figure~\ref{fig:MZ} shows that even in the M-Z relation, some low metallicity outliers remain among the XMP galaxies selected by \citet{McQuinn_2020}. One possibility to explain the objects that remain outliers in both the L-Z and M-Z relations is that the measured metallicity is not representative of the stellar mass of the underlying galaxy.  This lower metallicity is likely attributed to external events.
For example, this can happen if the \hii\ region results from SF in a pocket of pristine gas acquired from the circumgalactic medium \citep{Ekta_2010}, or if it is the result of a minor merger with a galaxy/component that has a lower metallicity \citep{Pascale_2021}. In these scenarios, the metallicity of the main galaxy would be underestimated, and the object would be an outlier in both L-Z and M-Z relations. 

\begin{figure}[h]
\includegraphics[width=0.99\linewidth]{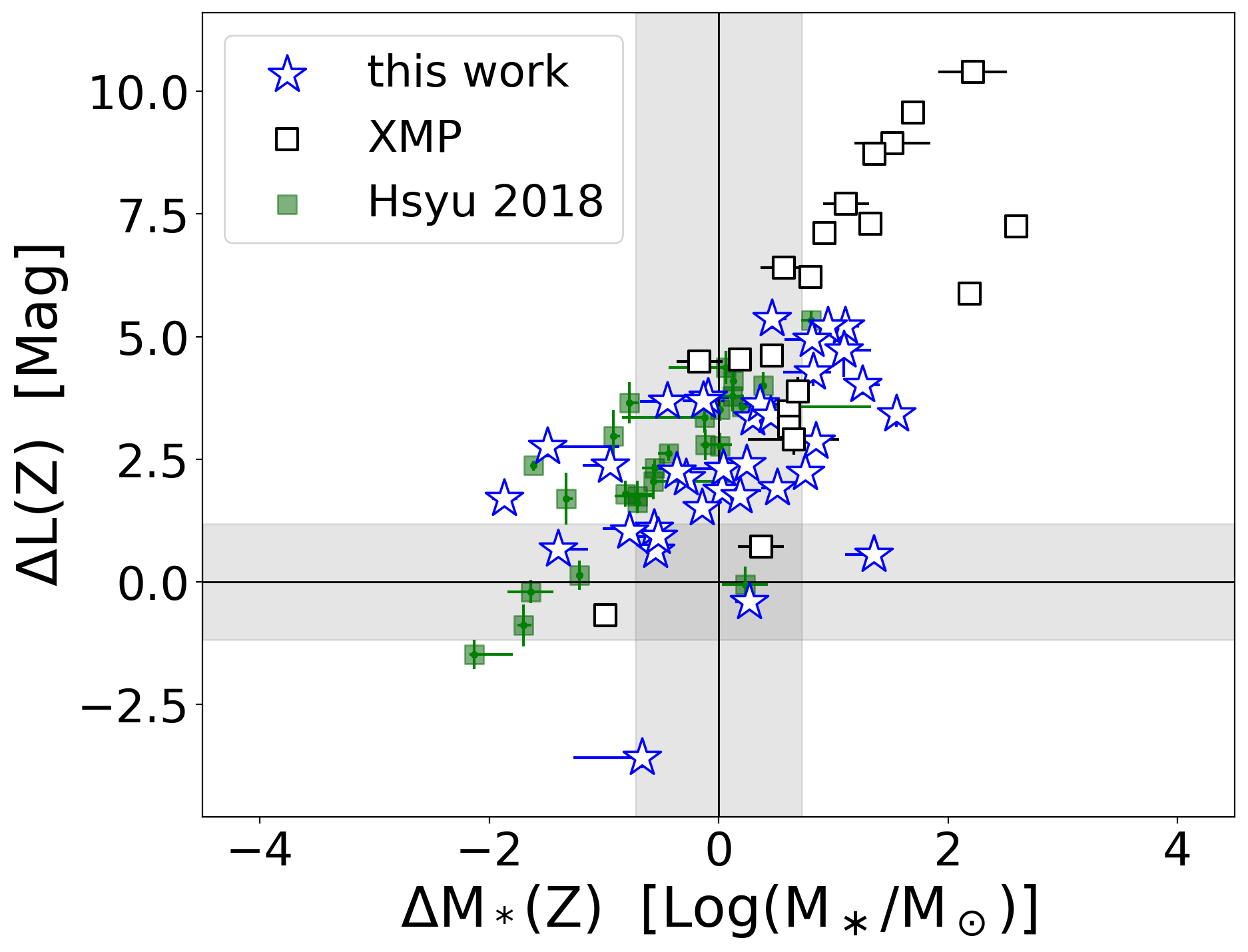}
\caption{  The horizontal offsets of the L-Z relation vs.  the M-Z relation.  The gray shaded areas are the 1~$\sigma$ scatter of the L-Z and M-Z relation. \label{fig:dMZLZ}}
\end{figure}

In an attempt to identify galaxies with possible low-metallicity inflows, we show in Figure~\ref{fig:dMZLZ} the luminosity and stellar mass offsets of the galaxies, and we define the offsets as the horizontal distance of the data from the M-Z and L-Z relations defined by the LVL galaxies.  
Most of the DES BCDs and the Hsyu et al.'s sample have small scatter from the M-Z relation, with $\Delta$M$_\ast$(Z)$<$1 $\sigma$ of the M-Z relaion, and larger scatter from the L-Z relation, with $\Delta$L(Z)$\sim$2 $\sigma$ of the L-Z relation.
For the XMP galaxies there appears to be a correlation between  $\Delta$L(Z) and $\Delta$M$_\ast$(Z), with most of the XMP galaxies located on the upper right side of the plot.  This result suggests that the ``extreme metal-poor" nature of these objects is a consequence of pristine gas inflow.  Although not as significant as for the XMP galaxies (which are extreme by definition), the offsets for the BCDs appear to correlate in a similar way to the XMP galaxies. This suggests that pristine inflows may in fact be very common, as required by galaxy formation models \citep{Dave_2012, Ma_2016}.

\subsection{Mass-Metallicity-Star Formation Rate Relation}

The fundamental metallicity relation (FMR) describes a second order dependence 
 of the M-Z relation on the SFR. Specifically, it is observed  that for a given stellar mass, the lower the metallicity, the higher the star formation rate of the galaxy.  
Many theoretical models  reproduce the observed relation as the result of the interplay among various effects: pristine gas brought in with the inflow fueling star-formation, enriched elements being expelled by  outflows, and the varying amount of gas available for star formation.  The dependence on the SFR is likely a reflection of the dependence on the gas fraction and the efficiency of the baryonic cycle \citep{Bothwell_2016a, Bothwell_2016b}.

Note that the SFR dependence on the fundamental metallicity relation is different from how the SFR affects the L-Z relation. The latter is biased by the SFR since the relation is determined from the luminosity of a single broad band filter.  

The FMR in \citet{Curti_2020} is derived using SDSS galaxies with stellar mass 8 $<$ log(M$_\ast$) $<$ 12, given as 
\begin{equation}
\label{eq:FMR_s}
Z_{FMR}= Z_0 + \frac{\gamma}{\beta}\text{log}(1 + (\frac{\text{M$_\ast$}}{M_0(\text{SFR})})^{-\beta}),
\end{equation}
where Z$_{FMR}$ is the oxygen abundance 12+\text{log(O/H)}, Z$_0$ = 8.779 $\pm$0.005, $\gamma$ = 0.31$\pm$0.01, $\beta$ = 2.1$\pm$0.4, log($M_0$(SFR)) = m$_0$ + m$_1$log(SFR), with  m$_0$= 10.11$\pm$0.03,  m$_1$ = 0.56$\pm$0.01.  The m$_0$ and Z$_0$ are the scale of the turnover mass and metallicity at the high mass end where the M-Z relation begins to flatten out. The dispersion in 12+log(O/H) is 0.054 dex, but increase toward low mass and high SFR end \citep[see also ][]{Hirschauer_2018}.  At the low mass end where M$_\ast << $m$_0$, this equation reduces to a linear relation: 

\begin{equation}
\begin{aligned}
\label{eq:FMR_p}
Z_{\text{FMR}} =& (0.31 \pm 0.01) \times \text{log(M$_\ast$)} + ( 5.65 \pm 0.01) \\
         &-(0.17 \pm 0.01) \times \text{log(SFR)}.
\end{aligned}
\end{equation}
Note that \citet{Curti_2020}  also provide the M-Z relation at the low-mass end, with $\gamma$ = 0.28$\pm$0.02, which agrees well with the M-Z relation found with the LVL galaxies in \citet{Berg_2012}. 

\begin{figure}[h]
\includegraphics[width=0.99\linewidth]{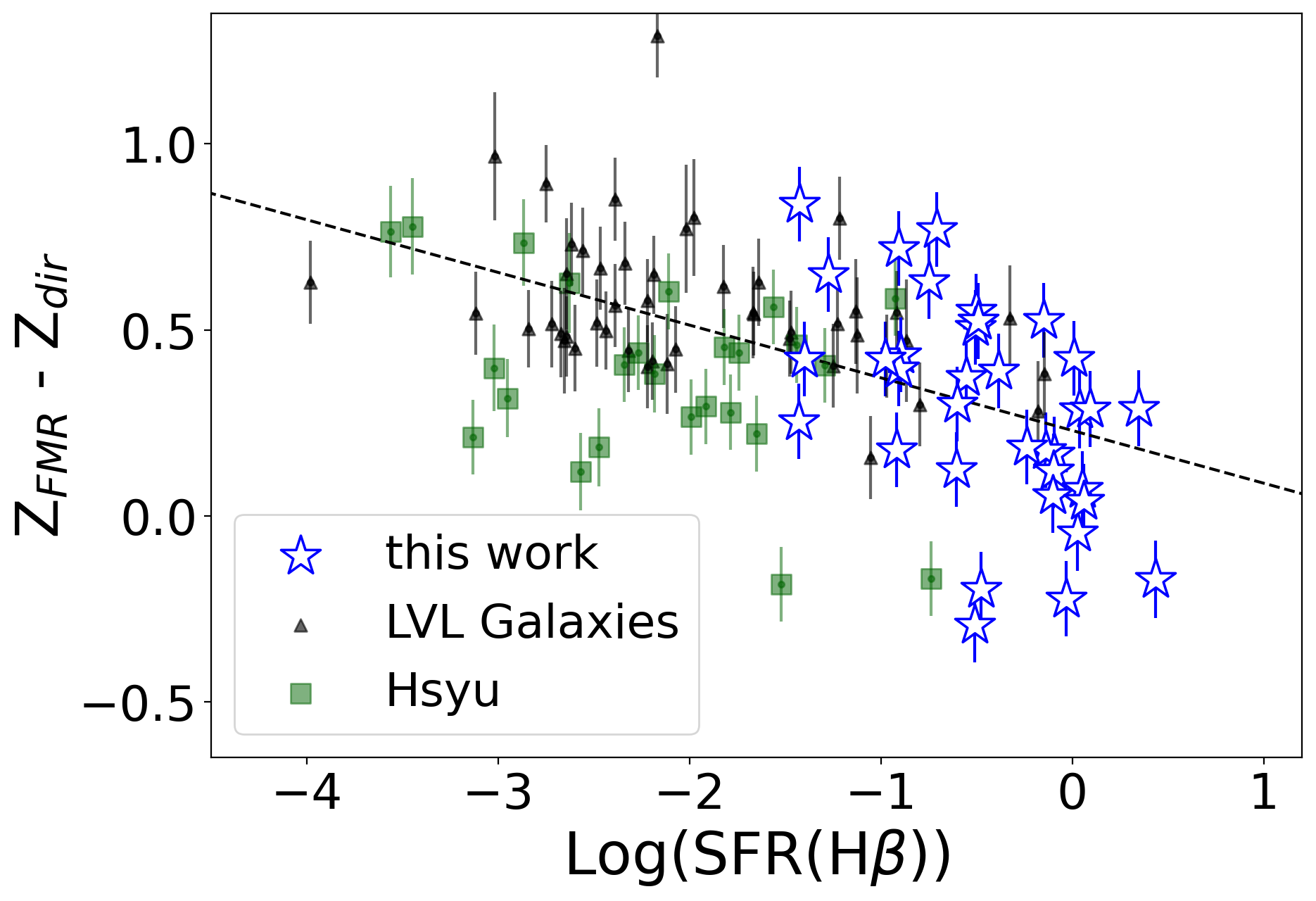}
\caption{ $Z_{\text{FMR}}$ - Z$_{dir}$ vs Log(SFR(\hb)).  The dash line is the best linear fit to the equation~(\ref{eqn:FMR_dZ}). } \label{fig:fmr_dZ}
\end{figure}

We compared the FMR derived for the  low mass objects with our DES galaxies, but find no significant second-order dependence of the metallicity on the SFR.  Objects with 12-log(O/H) $<$ 8.0 do not fall on the FMR surface described in equation~(\ref{eq:FMR_s}).  
When calculating the metallicity expected from the FMR relation for the stellar mass and SFR of each galaxy, we find that the metallicity residuals $\Delta Z = Z_{\text{FMR}} - Z_{dir}$ correlates with the SFR. This is shown in Figure~\ref{fig:fmr_dZ}, where Z$_{dir}$ is the metallicity measured in Section~\ref{sec:direct} using the direct method.  
The mean of the metallicity residuals $\Delta Z$ is 0.45 dex, greater than the uncertainty of $\sim$0.25 at the low mass end of FMR \citep[see ][ Figure 11]{Curti_2020}. 
Intriguingly, the metallicity residuals are actually larger for the lower SFR galaxies, while these lower SFR galaxies have the sSFR closer to the main-sequence galaxies. 
We fit a linear correlation to the residual $\Delta Z$ as a function of the Log(SFR):

\begin{equation} 
\label{eqn:FMR_dZ}
\Delta Z = (-0.14\pm0.02)  \times \text{log(SFR)} + 0.23 \pm 0.04. 
\end{equation}

We can use the relation above, to update the FMR in equation~\ref{eq:FMR_p}, and see that the dependence on SFR decreases:

\begin{equation} \label{eq:new_FMR}
\begin{aligned}
Z_{\text{FMR}} =& Z_{dir} + \Delta Z \\
        =&  (0.31 \pm 0.01) \times \text{log(M$_\ast$)} + ( 5.65 \pm 0.01) \\
        &-(0.17 \pm 0.01) \times \text{log(SFR)}, \\
Z_{dir} =& Z_{\text{FMR}}  - \Delta Z \\
=& (0.31 \pm 0.01) \times \text{log(M$_\ast$)} +  ( 5.42 \pm 0.04) \\ 
&- (0.03\pm0.02)  \times \text{log(SFR)}.
\end{aligned}
\end{equation}
{\bf In the updated equation~\ref{eq:new_FMR} the SFR dependence is weak enough that it can be neglected, and the resulting the M-Z relation is almost identical to the one of  \citet{Berg_2012}. Therefore the dispersion in the M-Z relation of the BCDs cannot be tightened by the star formation rate, and other factors may be driving it.}

The FMR plane is the consequence of “main sequence” galaxies reaching equilibrium between gas inflow, outflow, and star formation.  The extrapolation of the FMR plane cannot be taken literally, but what is the applicable lowest mass for this equilibrium model?  With bursty star formation histories and lower gravity, dwarf galaxies such as our BCDs can be easily driven away from the equilibrium between gas flows and star formation. However, regardless of the burstiness, the LVL galaxies, having similar sSFR to those of the “main sequence” galaxies, do not follow the FMR plane either. \citet{Llerena_2023} also find that lower sSFR galaxies have greater metallicity residuals from a $z\sim3$ SFG sample.  Detailed mechanisms of the balance between inflows and outflows may be at play in regulating the position of low mass galaxies in the M$_\ast$-SFR-Z space. For example, \citet{Magrini_2012} considers a multi-phase interstellar medium model, and shows that compact regions with dense gas have distinct properties from a diffuse region. The scatter of the M-Z relation may result from galaxies with different gas density. {\bf  \citet{Forbes_2014} shows that a large intrinsic scatter of the mass loading factor (the ratio between the ejected mass from galatic winds and the galaxies' SFR) will weaken the anticorrelation between metallicity and SFR, at a fixed mass. The variation among mass loading factors in low mass galaxies can be enhanced by galactic winds generated from supernova explosions. These galactic winds are metal-enriched \citep{Vader_1986, Vader_1987}, and may escape (blowout) or even remove the ISM from the galactic halo (blow-away) given their shallow gravitational potential \citep{MacLow_1999}. For example, \citet{McQuinn_2015b} finds the low metallicity dwarf galaxy Leo~P held on to only 5$\%$ of the oxygen atoms, the other 95$\%$ was presumably lost to the galaxy halo or the intergalactic medium.  The oxygen retained in Leo P is much lower than estimates of 20$\%$–25$\%$ of metals retained in more massive galaxies \citep[$10^9 M_\odot < M_* < 10^{11.5} M_\odot$,][]{Peeples_2014}. }

\subsection{Strong Line Calibration}\label{sec:discuss_SL}

\begin{figure}[h]
\includegraphics[width=0.99\linewidth]{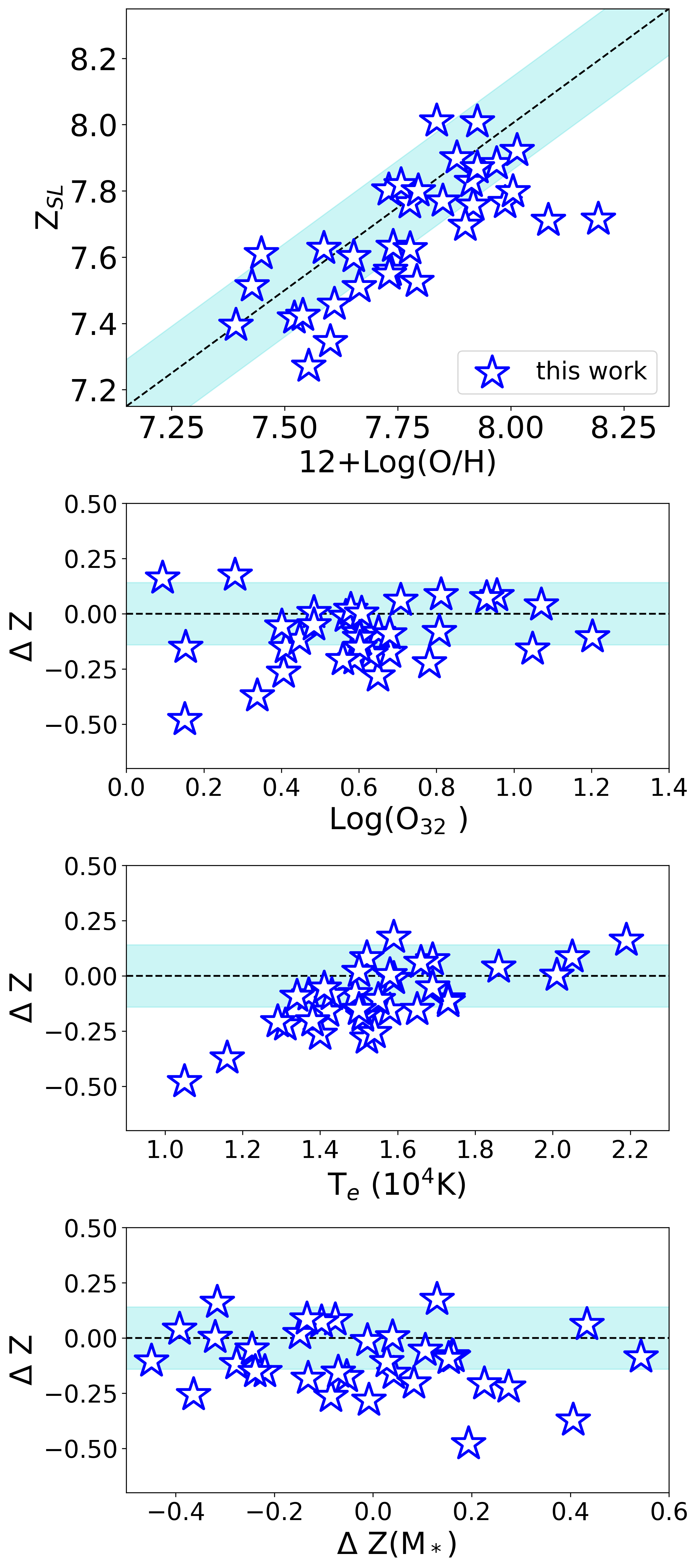}
\caption{First panel: 12 + log(O/H) calculated via the direct method (Z$_{dir}$) vs. Strong Line Calibration (Z$_{SL}$) in \citet{Jiang_2019}. The dashed line is the identical line. The light blue shadows are $\pm$0.1 dex as uncertainty. Second panel: The metallicity difference of the two methods $\Delta$Z = Z$_{SL}$-Z$_{dir}$ vs. Log(O$_{32}$). The light blue shadow indicates $\pm$0.16 dex as uncertainty in the strong line calibration as per \citet{Jiang_2019}. Third panel:  $\Delta$Z vs. T$_e$/10$^4$K.  Fourth panel: $\Delta$Z vs. Metallicity offset in the M-Z relation. \label{fig:slc_J}}
\end{figure}

Here we explore the reliability of strong line calibrations described in Section~\ref{sec:SL} with the BCDs presented in this work. 
Note that while there are many strong line calibrations \citep[e.g][]{Maiolino_2008, Izotov_2021}, we choose \citet{Jiang_2019} specifically because it is calibrated on green pea galaxies and is the most directly applicable calibration available for the metallicities of our sample.
Figure~\ref{fig:slc_J} shows the comparison of the metallicity derived from the direct method (Z$_{dir}$) and the strong line calibration (Z$_{SL}$) in \citet{Jiang_2019}. For most objects, the metallicity calculated from the strong line method is slightly lower, within 0.2 dex, compared to the values computed from the direct method. The mean offset is $\Delta$Z = Z$_{SL}$ - Z$_{dir}=  0.10 \pm 0.16$ dex.  
The second panel of Figure~\ref{fig:slc_J} shows $\Delta$Z versus the O$_{32}$ ratio. The third panel of Figure~\ref{fig:slc_J} shows $\Delta$Z versus T$_e$, and the fourth panel shows $\Delta$Z versus the metallicity offset in the M-Z relation.

As for the photoionization-parameter-sensitive line ratio O$_{32}$, the scatter $\Delta$Z is greater when Log(O$_{32}$) $<$ 0.5, consistent with the result in \citet{Jiang_2019}. The observed wavelength of the [\oii]$\lambda$ 3727 emission line is at the blue end of our spectra, where the uncertainty of the flux calibration is greater.  An overestimation on [\oii]$\lambda$ 3727 would lead to lower Log(O$_{32}$) and higher metallicity Z$_{dir}$.  A low O$_{32}$ line ratio, if measured correctly, represents a lower temperature and a fainter [\oiii]$\lambda$ 4363 line, which increases the uncertainty in the direct method.  This is consistent with the tighter correlation between the $\Delta$Z and the electron temperature T$_e$, as T$_e$ is determined without the [\oii] emission.  
The metallicity of the outliers in the M-Z relation is caused by the dilution of the infall pristine gas rather than the whole galaxy. This dilution does not affect the estimated metallicity. As expected, we do not observe a significant correlation between the $\Delta$Z and the metallicity offset from the M-Z relation ($\Delta$Z(M$_\ast$) =  Z$_{dir}$ - Z(M$_\ast$)).

\section{Conclusions}\label{sec:conclusion}

We present a new selection of 358 blue compact dwarfs from 5,000 deg$^2$ in the DES imaging survey, and the spectroscopic confirmation of a sub-sample of 64 galaxies using the CTIO COSMOS 4m telescope. The mean redshift of the confirmed BCDs is 0.04.

For the spectroscopic subsample, we estimate the galaxies' luminosities, stellar masses,  and star formation rates from the combination of imaging and spectroscopic data. For 34 galaxies with deep spectroscopic follow-up we measure the metallicity via the direct T$_e$ method using the auroral [\oiii]$\lambda$ 4363 emission line. The metallicity of our objects ranges from 7.37 $\leq$ 12+log(O/H) $\leq$ 8.16, with a mean of 12+log(O/H)= 7.8.  The SFRs are in the range of 0.03 to 3 M$_\odot$ yr$^{-1}$, with stellar masses between 10$^7$ to 10$^8$ M$_\odot$.  The sSFR of the BCDs are around 10$^{-9}$ to 10$^{-7}$ yr$^{-1}$.  

We compare the position of our BCDs with the M-Z and L-Z relation derived from the Local Volume Legacy sample.
We find that in the L-Z plane, the DES BCDs fall below the LVL L-Z relation but are above the most extreme metal-poor galaxies.  In contrast, in the M-Z plane, the DES BCDs agree with the local M-Z relation, and the scatter around the M-Z relation is smaller  than the scatter around the L-Z relation. We identify a correlation between the offsets from the M-Z and L-Z relation that we suggest is due to the contribution of metal-poor inflows. 

Finally, we explore the validity of the mass-metallicity-SFR fundamental plane in the mass range probed by our galaxies ($<10^{8}$M$_{\odot}$). This is the largest sample of  galaxies in this mass range with accurate metallicity that can be used for such a study. 
Our results show that low mass  star forming galaxies do not follow the extrapolation of the fundamental plane. This result suggests that mechanisms other than the balance between inflows and outflows may be at play in regulating the position of low mass galaxies in the M-SFR-Z space.
Recent JWST spectroscopic results show that $z>7$ galaxies behave in a similar way as our sample of  BCDs: they are consistent with the $z=0$ M-Z relation, while deviating from the mass-metallicity-SFR fundamental plane \citep{Curti_2022}. Understanding this common behavior at the lowest masses will shed light on the metal enrichment of galaxies in the early universe.   


\section*{Acknowledgements}
All the {\it GALEX} data used in this paper can be found in MAST: \dataset[10.17909/T9H59D]{https://archive.stsci.edu/missions-and-data/galex}.
This research uses services or data provided by the Astro Data Lab at NSF’s NOIRLab. NOIRLab is operated by the Association of Universities for Research in Astronomy (AURA), Inc. under a cooperative agreement with the National Science Foundation.  MH is fellow of the Knut $\&$ Alice Wallenberg Foundation.  


\appendix

\section{SQL Query}\label{apd:SQL}

\texttt{ SELECT coadd$\_$object$\_$id, ra, dec,  snr$\_$g, snr$\_$r, snr$\_$i, snr$\_$z,  mag$\_$auto$\_$g, mag$\_$auto$\_$r, mag$\_$auto$\_$i, mag$\_$auto$\_$z, 
magerr$\_$auto$\_$g, magerr$\_$auto$\_$r, magerr$\_$auto$\_$i, magerr$\_$auto$\_$z, ebv$\_$sfd98 \\
FROM des$\_$dr1.galaxies  \\
WHERE snr$\_$g > 5  \\
AND snr$\_$r > 5 \\
AND snr$\_$i > 5 \\
AND snr$\_$z > 3 \\
AND (mag$\_$auto$\_$g - mag$\_$auto$\_$r) BETWEEN -0.2 and 0.2 \\
AND (mag$\_$auto$\_$r - mag$\_$auto$\_$i) BETWEEN -0.7 and -0.1 \\
AND (mag$\_$auto$\_$i - mag$\_$auto$\_$z) < 0.1  \\
AND (mag$\_$auto$\_$i - mag$\_$auto$\_$z) > -0.4 - (2*magerr$\_$auto$\_$z) \\
AND mag$\_$auto$\_$g > 22 \\
AND mag$\_$auto$\_$r > 22 \\
AND flags$\_$g = 0  \\
AND flags$\_$r = 0  \\
AND flags$\_$i = 0  \\
}

\section{Slit Loss Correction}\label{apd:slitloss}
In this section we describe the dispersion correction we apply on the 2020 observation run. 
Following the equation (1) and (2) in \citet{Filippenko_1982}, we first calculate the refractive index $n(\lambda)_{T,P}$, with temperature T $\simeq$ 9$^{\circ}$C and P $\simeq$ 600 mm Hg, according to our observation condition. 
The  atmospheric differential refraction relative to $\lambda$ = 5000\AA\ is 
\begin{equation}
    \Delta R(\lambda) \equiv R(\lambda) - R(5000\AA) 
    \simeq 20625 ( n(\lambda) - n(5000\AA) ) \text{tan}z, 
\end{equation}
where $z$ is the zenith angle. 

We assume the target point spread function as a Gaussian profile, with the seeing being the diameter (2$\sigma$) of the profile, 
\begin{equation}
    \mu(x, y)= \frac{1}{2\pi\sigma^2} e^{-(x^2+y^2)/2\sigma^2}. 
\end{equation}

The amount of light that goes into the slit with slit width 2a and slit length 2b arc second is
\begin{equation}
   I(\lambda) = \int^a_{-a} \int^b_{-b} \mu(x, y) dx~dy.  
\end{equation}
The refraction $\Delta$R($\lambda$) moves the light profile away from the center of the slit, 
\begin{equation}
    \Delta x \simeq \Delta R sin(\theta);~ \Delta y \simeq \Delta R cos(\theta),
\end{equation}
where $\theta$ is the difference between the slit position angle and the parallactic angle. 

With the refraction, the amount of light that goes into the slit is 
\begin{equation}
    I^{\prime}(\lambda) = \frac{1}{4} ( \int^{(a-\Delta x)}_{-(a-\Delta x)} \mu(x) dx + \int^{(a+\Delta x)}_{-(a+\Delta x)} \mu(x) dx  )\times
    ( \int^{(b-\Delta y)}_{-(b-\Delta y)} \mu(y) dy + \int^{(b+\Delta y)}_{-(b+\Delta y)} \mu(y) dy  ).
\end{equation}
$I^{\prime}(\lambda)$ is the observed flux density, and $I(\lambda)$ is the flux density that result to intrinsic emission line ratios. We define the flux recovery factor $\epsilon(\lambda)$ as $\epsilon(\lambda) \equiv I(\lambda)/I^{\prime}(\lambda)$, and multiple the factors along the spectra.

\bibliography{sample631}{}
\bibliographystyle{aasjournal}

\end{document}